\documentclass[preprint,12pt,3p]{elsarticle}
\usepackage[utf8]{inputenc}
\usepackage{amsmath}
\usepackage{amssymb}
\usepackage{graphicx}
\usepackage{xcolor} 
\usepackage{float} 
\usepackage{siunitx}
\usepackage{stmaryrd}
\usepackage{url}
\usepackage[normalem]{ulem}
\usepackage[colorinlistoftodos]{todonotes}
\usepackage[pdfencoding=auto]{hyperref}
\usepackage{booktabs}

\newcommand{\eqn}[1]{\begin{align}#1\end{align}}
\newcommand{\bs}[1]{\boldsymbol{#1}}

\newcommand{\pare}[1]{\left( #1 \right) }
\newcommand{\corchete}[1]{\left[ #1 \right]}
\newcommand{\fr}[2]{\frac{#1}{#2}}
\newcommand{\wtil}[1]{\widetilde{#1}}
\newcommand{\mc}[1]{\mathcal{#1}}

\newcommand{\tex}[1]{\mbox{\scriptsize{#1}}}

\newcommand{\what}[1]{\widehat{#1}}

\definecolor{darkgreen}{rgb}{0, 0.5, 0.05}

\newcommand{\deleted}[1]{}
\newcommand{\refone}[1]{{#1}}
\newcommand{\refthree}[1]{{#1}}
\newcommand{\new}[1]{{#1}}
\newcommand{\crossout}[1]{}

\def\dt{\Delta t}
\def\dd{\mathrm{d}}  

\def\bna{\bs{\nabla}}
\def\bDl{\Delta \bs{l}}

\def\bC{\bs{C}}

\def\bx{\bs{x}}
\def\bI{\bs{I}}
\def\bK{\bs{K}}
\def\bp{\bs{p}}
\def\bP{\bs{P}}
\def\bM{\bs{M}}
\def\bR{\bs{R}}
\def\bomega{\bs{\omega}}

\def\bt{\bs{t}}
\def\bT{\bs{T}}
\def\bt{\bs{t}}
\def\btau{\bs \tau}

\def\bb{\bs{b}}
\def\bB{\bs{B}}
\def\be{\bs{e}}
\def\bu{\bs{u}}
\def\bv{\bs{v}}
\def\bU{\bs{U}}
\def\bl{\bs l}
\def\blambda{\bs{\lambda}}
\def\br{\bs{r}}
\def\bF{\bs{F}}
\def\bbf{\bs{f}}
\def\bphi{\bs{\phi}}
\def\bq{\bs{q}}
\def\bQ{\bs{Q}}
\def\bdx{\bs{\delta x}}
\def\bdq{\bs{\delta q}}
\def\bdth{\bs{\delta \theta}}
\def\bg{\bs{g}}
\def\bG{\bs{G}}
\def\bA{\bs{A}}
\def\bN{\bs{N}}
\def\bJ{\bs{J}}
\def\by{\bs{y}}
\def\bz{\bs{z}}

\def\bzero{\bs{0}}
\def\btheta{\bs \theta}
\def\bgamma{\bs \gamma}

\def\mcA{\mc{A}}
\def\mcB{\mc{B}}
\def\mcL{\mc{L}}

\journal{}

\begin{document}

\begin{frontmatter}

\title{A numerical method for suspensions of articulated bodies in viscous flows}

\author[inst1]{Florencio Balboa Usabiaga}

\affiliation[inst1]{organization={BCAM - Basque Center for Applied Mathematics},
            city={Mazarredo 14, Bilbao},
            postcode={E48009}, 
            country={Basque Country - Spain}}

\author[inst2]{Blaise Delmotte}

\affiliation[inst2]{organization={LadHyX, CNRS, Ecole Polytechnique, Institut Polytechnique de Paris},
            city={Palaiseau},
            postcode={91120}, 
            country={France}}

\begin{abstract}

An articulated body is defined as a finite number of rigid bodies connected by a set of arbitrary constraints that limit the relative motion between pairs of bodies. Such a general definition encompasses a wide variety of situations in the microscopic world, from bacteria to synthetic micro-swimmers, but it is also encountered when discretizing inextensible bodies, such as filaments or membranes.
In this work we consider hybrid articulated bodies, i.e.\ constituted of both linear chains, such as filaments, and closed-loop chains, such as membranes. 
Simulating suspensions of such articulated bodies requires to solve the hydrodynamic interactions between large collections of objects of arbitrary shape while satisfying the multiple constraints that connect them. 
Two main challenges arise in this task: limiting the cost of the hydrodynamic solves, and enforcing the constraints within machine precision at each time-step.
To address these challenges  we propose a formalism that combines the body mobility problem in Stokes flow with a velocity formulation of the constraints, resulting in a mixed mobility-resistance problem.
While resistance problems are known to scale poorly with the particle number, our preconditioned iterative solver is not sensitive to the system size,
therefore allowing to study large suspensions with quasilinear computational cost.
Additionally, constraint violations, e.g.\ due to discrete time-integration errors, are prevented  by correcting the particles' positions and orientations at the end of each time-step. Our correction procedure, based on a nonlinear minimisation algorithm, has negligible computational cost and preserves the accuracy of the time-integration scheme. 
The versatility of our method allows to study a plethora of articulated systems within a unified framework.
We showcase its robustness and scalability by exploring the locomotion modes of a model microswimmer inspired by the diatom colony \textit{Bacillaria Paxillifer}, and by simulating large suspensions of bacteria interacting near a no-slip boundary. Finally, we provide a Python implementation of our framework in a collaborative publicly available code, where the user can prescribe a set of constraints through a   single input file to study a wide spectrum of applications involving suspensions of articulated bodies.
\end{abstract}




\end{frontmatter}

\tableofcontents

\section{Introduction}
An articulated body consists of a set of rigid bodies connected together by constraints that limit the relative motion between pairs of bodies. 
Articulated bodies are ubiquitous in microscopic systems.
Bacteria, the dominant prokaryotic microorganisms, self-propel with one or several helical flagella attached to their rigid head. The head and flagella  are articulated together by an inextensible hook connected to a rotary motor to achieve  self-propulsion \cite{Berg1973,Block1989,Trachtenberg1992}. In the oceans and waterways, small unicellar organisms, called diatoms, can assemble into  colonies with complex articulations. The species \textit{Bacillaria Paxillifer} forms  colonies of  stacked rectangular cells that slide along each other while remaining parallel. Their intriguing coordinated motion  leads to beautiful and nontrivial trajectories at the scale of the colony  \cite{Muller1782,Kapinga1992,Gordon2016}.
Inspired by Nature, scientists have designed articulated systems that can achieve locomotion. Artificial microswimmers use a plethora of swimming gaits such as the undulation of hinged rigid segments, or colloidal beads linked by DNA,  under the action of a magnetic field \cite{Dreyfus2005,Jang2015,Liao2019},  articulated arms moving in a non-reciprocal manner to mimic elementary propulsion mechanisms, such as the well-known Purcell's swimmer \cite{Purcell1977,Becker2003}, or more elaborate strategies such as the four-arm breaststroke of the \textit{Copepod} zooplankton \cite{Bonnard2018,Bettiol2018}.
Beyond microswimmers, small articulated systems  are investigated for the design of smart composite particles and new functional materials. Using  DNA functionalized colloids and emulsions respectively,  Sacanna et \textit{al.}\ \cite{Sacanna2010} and  McMullen et \textit{al.}\ \cite{McMullen2018} both showed that polymeric freely-jointed molecules can be assembled at the micron scale.

Joint articulations are also encountered  when discretizing inextensible objects, such as fibers or membranes, with numerical methods. As shown in the literature \cite{Ross1997,Delmotte2015b,Schoeller2021}, the inextensibility condition along a fiber centerline can be discretized as a set of ball-and-socket joints, i.e.\ only allowing  rotation, between discrete degrees of freedom. 
Regardless of their classification, constraints in articulated bodies can either form open chains, such as filaments,  closed-loop chains, such as membranes, or a combination of both.

Simulating suspensions of such a wide variety of articulated bodies requires to solve the hydrodynamic interactions between large collections of rigid objects of arbitrary shape while satisfying multiple constraints.
Developing efficient methods to solve for the constrained motion of immersed objects is an active subject of research. One of the main challenges in this task is to build  scalable solvers for the hydrodynamic and constraint equations. 
To solve the hydrodynamic problem, numerical methods, such as the Boundary Element method \cite{Pozrikidis1992,Montenegro2015} or the rigid multiblob method \cite{Swan2016,Usabiaga2016}, discretize  particles of arbitrary shape with discrete degrees of freedom, or markers, constrained to move as a single rigid body.
The hydrodynamic interactions between discrete degrees of freedom are materialized by a dense  mobility matrix.  The mobility matrix is a position-dependent linear operator that relates the forces applied on the markers to their velocities. 
Thanks to  key advances made in the past decade, the action of the mobility matrix on a vector, a bottleneck in terms of computational cost, can be computed in a fast and scalable way.
These  approaches  rely on  fast  summation techniques such as the FMM \cite{Greengard1987,Tornberg2008,Yan2020} or the FFT, and include Ewald summation \cite{Wang2016,Fiore2017}, Accelerated and Fast Stokesian Dynamics \cite{Sierou2001,Fiore2019}, in addition to the Immersed Boundary \cite{Peskin2002} and similar methods, such as the Force Coupling Method \cite{Maxey2001,Lomholt2003}, that make use of fast, matrix-free, solvers.

Following classical mechanics, the coupling between the constraints and the hydrodynamic part is achieved through a set of Lagrange multipliers that are translated into constraint forces on the bodies through the Jacobian of the constraint equations.
The way these Lagrange multipliers are determined depends on the constraint formulation.
When the constraints are holonomic, i.e.\ they depend on the particle positions, orientations and time,  finding the Lagrange multipliers requires solving a nonlinear system in the particle positions and orientations \cite{Goldstein2002}.
In this case, the constraint forces are given by the  product between the vector of Lagrange multipliers and the Jacobian of the constraints with respect to the particle's positions and orientations.
If the constraint are nonholonomic, linear and exact, i.e.\ they  depend linearly on the particle velocities and are exact differential equations with respect to time, then one can use the  linearity of Stokes equations between forces and velocities to find the Lagrange multipliers directly. When inverting the resulting linear system, one  computes the \textit{constraint resistance matrix} that involves the mobility matrix between markers \cite{Delmotte2015b}. The corresponding constraint forces are given by the  product between the vector of Lagrange multipliers and the Jacobian of the constraints with respect to the particles' velocities. 

Up to now, numerical methods for suspensions of articulated bodies have been specifically devoted to the modeling of active and passive inextensible filaments as linear chains of freely-jointed spherical bodies. 
Most of them are built upon the velocity (i.e.\ nonholonomic) formulation  of the free-joint  inextensibility constraints and use analytical expressions of the mobility matrix between the spherical  bodies  distributed along the filament centerline \cite{Yamamoto1993,Ross1997,Lindstrom2007,Schmid2000,Delmotte2015b}. Even though the velocity formulation combines nicely with the linearity of the mobility problem, these methods rely on direct dense linear algebra tools to find the set of Lagrange multipliers, and therefore scale badly with the number of fibers. 
Schoeller \textit{et al.} \cite{Schoeller2021} recently overcame  that limitation by using holonomic inextensibility constraints with an efficient nonlinear solver to simultaneously find the Lagrange multipliers and the implicit update of the sphere  positions and orientations. Their iterative scheme, combined with the Force Coupling Method \cite{Maxey2001,Lomholt2003} on a grid-based solver, allowed them to simulate large suspensions with up to 1000 filaments for a wide variety of applications \cite{Schoeller2018,Schoeller2021}. 
Though scalable to many particles, their method is specifically calibrated for linear chains of spherical bodies connected with free-joint constraints. 
Articulated bodies made of nonspherical units have been employed to study the dynamics of a single bacterium \cite{Shum2010,Shum2017} or parasite \cite{Walker2019}
with the Boundary Element Method (BEM).
However the constraint formulations used in these works are not generic, in the sense that they are specific to these micro-swimmers, and BEM can bear very small numbers of particles due to high computational costs.

Another drawback of the nonholonomic formulation of the constraints is that, even if the body velocities satisfy the constraints exactly, the local truncation errors due to discrete  time-integrators  for the body equations of motion, can accumulate and thus violate the position and orientation constraints. Various workarounds have been proposed such as recursive time-step reductions \cite{Lindstrom2007}, which can significantly increase the computational cost, or successive position adjustments \cite{Doi1989}.

In this study, we provide a robust, comprehensive and scalable framework to handle large collections of articulated bodies composed of particles with arbitrary shape connected in an arbitrary fashion. 
The mobility problem between rigid bodies of arbitrary shape is discretized with the rigid multiblob model, which is formulated as a symmetric saddle system involving the mobility matrix between markers \cite{Usabiaga2016}. Expanding on our previous work \cite{Delmotte2015b}, we combine the mobility problem  with a velocity formulation of the constraints, resulting in a mixed mobility-resistance problem for the unknown body velocities and constraint Lagrange multipliers. 
Instead of directly solving the corresponding linear system, we use an iterative method with a block-diagonal preconditioner. Since iterative solvers only require computing the action of the mobility matrix onto a vector, our framework works  with any  fast method.
We demonstrate the efficiency of the preconditioner on various configurations involving articulated bodies with open and closed loops, such as deformable filaments and shells.
While resistance problems are known to scale poorly with the particle number, our preconditioned iterative solver is not sensitive to the system size, therefore allowing to simulate large suspensions with quasilinear computational cost. 
The number of degrees of freedom in the system is then reduced by extending  the well-known robot-arm model to the more general case of hybrid chains, so that only one reference position and the bodies orientations are needed to track and uniquely reconstruct an articulated body. Our \refone{new} reconstruction scheme avoids  constraint violations for open chains but does not \new{fully} prevent them for closed loops or hybrid chains.
In order to prevent any constraint violation, e.g.\ due to  time discretization errors,  the particle positions and orientations are corrected  at the end of each time-step. Our correction procedure, based on a nonlinear minimisation algorithm, is negligible in terms of computational cost and preserves the accuracy of the time-integration scheme. 
Besides scaling and convergence tests on shells and filaments, we illustrate the versatility of the method with various simulations of active systems.
We first  explore the  locomotion modes of a model microswimmer made of sliding rigid rods, directly inspired by the diatom colony \textit{Bacillaria Paxillifer}. \refone{Very little, if nothing, is known about the swimming mechanisms of this species.  Our hydrodynamic simulations, seemingly the first ones of such microorganism,}  \refone{show} that, contrary to flagellar dynamics, the swimming speed and direction \refone{of the colony} depend nonmonotonically  on the wavelength of  \refone{its} deformation wave.
We also  study the swimming speed of a model bacterium, made of a rigid helical flagellum and spherical head, as a function of the helical wave number \refone{and compare our results with the theoretical work of Higdon \cite{Higdon1979}}. Finally we simulate the collective
dynamics of \refone{large, fully resolved,}  bacterial suspensions near a surface \refthree{at a scale that, to the best of our knowledge, had never been reached for such systems.}

Our framework  has been implemented in a collaborative publicly available code on GitHub (\url{https://github.com/stochasticHydroTools/RigidMultiblobsWall}).
Our tool can handle different populations of articulated bodies simultaneously where both passive and time-dependent constraints are directly read from a simple  input file.
Thanks to its simplicity and flexibility, the user can readily use it to study physical and biological systems involving large collections of articulated bodies.


\section{Model and formulation}
\label{sec:model}
\refone{In this section we first outline the  equations of the mobility problem that govern the motion of rigid bodies of arbitrary shape freely suspended  in a viscous fluid. Then we write the  equations of the position constraints that connect these bodies. Expanding on the work of some of us \cite{Delmotte2015b}, we derive the velocity form of these constraints, which we combine to the mobility problem to propose a new formulation of the \textit{constrained mobility problem}. Finally we show how to use our formulation with existing methods to solve the mobility problem, such as the rigid multiblob model \cite{Usabiaga2016}.}

\subsection{Continuum formulation for rigid bodies}
\label{sec:continuum}

We first introduce the formulation for a suspension of disconnected rigid bodies and then we generalize it to introduce constraints. 
Let $\{\mcB_p\}_{p=1}^M$ be a set of $M$ rigid bodies immersed in a Stokes flow.
The configuration of each rigid body $p$ is described by the location of a tracking point, $\bq_p$,
and its orientation represented by the unit quaternion $\btheta_p$, or in compact notation $\bx_p=\{\bq_p, \btheta_p\}$.
The linear and angular velocities of the tracking point are $\bu_p$ and $\bomega_p$.
The external force and torque applied to a rigid body are $\bbf_p$ and $\btau_p$.
We will use the concise notation $\bU_p=\{\bu_p, \bomega_p\}$ and $\bF_p=\{\bbf_p, \btau_p\}$ as well, while
 unscripted vectors will refer to the composite vector formed by the variables of all the bodies,
e.g.\ $\bU = \{\bU_p\}_{p=1}^M$.
The velocity and pressure, $\bv$ and $p$, of the fluid with viscosity $\eta$ are governed by the Stokes equations  \cite{Kim1991, Pozrikidis1992}
\eqn{
  \label{eq:Stokes}
  -\bna p + \eta \bna^2 \bv &= \bzero, \\
  \label{eq:div}
  \bna \cdot \bv &= 0,
}
while at the bodies surface the fluid obeys the no-slip condition \cite{Pozrikidis1992}
\eqn{
  \label{eq:no-slip-continuum}
  \bv(\br) = (\bs{\mc{K}} \bU)(\br) = \bu_p + \bomega_p \times (\br - \bq_p)\; \mbox{for }\br \in \partial \mcB_p,
}
where we have introduced the geometric operator $\bs{\mc{K}}$ that transforms rigid body velocities to surface velocities.
Since inertia does not play a role in Stokes flows the conservation of linear and angular momentum
reduces to the balance of force and torque. For every body $p$ the balance between hydrodynamic and external stresses is given by \cite{Pozrikidis1992}
\eqn{
  \label{eq:balanceF_continuum}
  \int_{\partial \mcB_p} \blambda(\br) \,\dd S_{\br} = \bbf_p, \\
  \label{eq:balanceT_continuum}
  \int_{\partial \mcB_p} (\br - \bq_p) \times \blambda(\br) \,\dd S_{\br} = \btau_p,
}
 where $-\blambda$ is the hydrodynamic traction exerted on the bodies by the fluid.
The adjoint of the geometric operator $\bs{\mc{K}}$ can be used to write the balance of force and torques for all bodies as  $\bs{\mc{K}}^T \blambda = \bF$.

Given the force and torque acting on the rigid bodies, the equations \eqref{eq:Stokes}-\eqref{eq:balanceT_continuum} can be solved
for the bodies velocities and the traction. Since the Stokes equations are linear we can write the velocity of $M$ rigid bodies as
\eqn{
  \label{eq:U_NF}
  \bU = \bN \bF,
}
where $\bN=\bN(\bx)$ is a $6M \times 6M$ mobility matrix that couple the forces and torques acting on the rigid bodies with their velocities.
Having $\bU$, the equations of motion can be integrated in time. 
Some care is necessary to integrate the quaternion equations, therefore we discuss briefly some of their main properties \cite{Delong2015b, Westwood2021}.
A unit quaternion, $\btheta=\corchete{s, \bp}$ with $s\in\mathbb{R}$ and $\bp \in \mathbb{R}^3$,  represents a rotation around a fixed axis: the finite rotation given by the vector $\bs{\gamma}$ has the  associated unit quaternion
\eqn{
  \btheta_{\bs{\gamma}} = \corchete{\cos(\gamma / 2),\, \sin(\gamma / 2) \bs{\gamma}/\gamma},
}
where $\gamma = \|\bs{\gamma}\|_2$. 
Unit quaternions can be combined by the quaternion multiplication 
\eqn{
  \label{eq:quaternion_mult}
  \btheta_3 = \btheta_2 \bullet \btheta_1 = \left[
      s_2 s_1 - \bp_2 \cdot \bp_1,\;
      s_2 \bp_1 + s_1 \bp_2 + \bp_2 \times \bp_1 \right].
}
Therefore, $\btheta_3$ represents the rotation obtained by the combination of a rotation $\btheta_1$ followed by a rotation $\btheta_2$.
This product rule allows to write the kinematic equations of motion as
\eqn{
  \label{eq:dq_dt}
  \fr{\dd \bq_p}{\dd t} =& \bu_p, \\
  \label{eq:dtheta_dt}
  \fr{\dd \btheta_p}{\dd t} =& \fr{1}{2}\corchete{0, \bomega_p} \bullet \btheta_p.
}


\subsection{Constrained mobility problem}
\label{sec:kin_const}
\begin{figure}
  \begin{center}
  \includegraphics[width=0.99 \columnwidth]{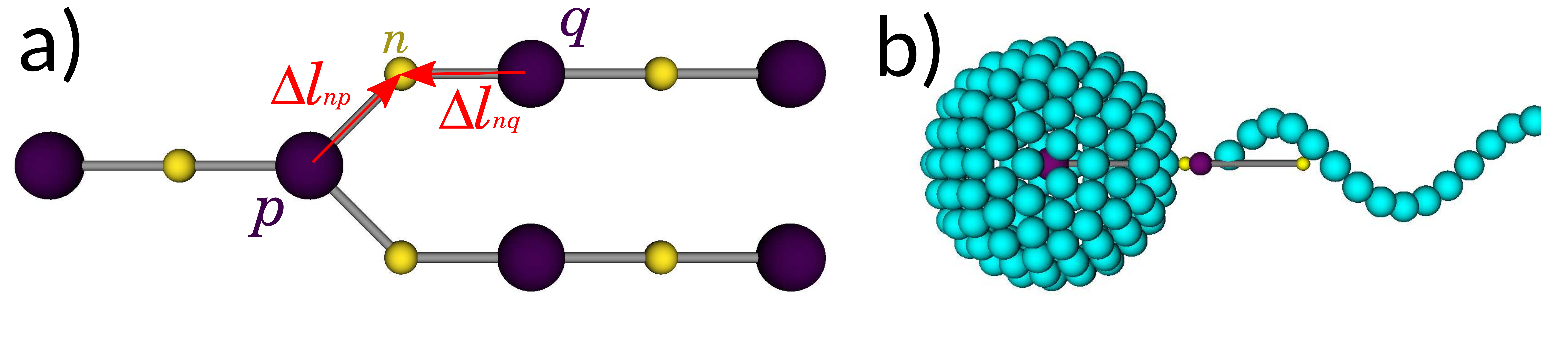}
  \caption{a) Example of a branched filament modeled as an articulated body formed by six spherical bodies (large spheres)
    connect by joints (small spheres).
    The links connecting two rigid bodies, $p$ and $q$, through the joint $n$ are labeled in red.
   \new{ b) Bacterium model, its body and flagellum are discretized with blobs (blue spheres) and the tracking points of the body and flagellum (dark spheres)
    are connected by two links through joints (small spheres).
    The links fix the flagellum attachment point and the axis of rotation, see Sec. \ref{sec:time_convergence}.}
  }
  \label{fig:sketch_joint}
  \end{center}
\end{figure}

Our constraint formulation is inspired by the robot-arm parametrization \cite{Featherstone1987} generalised to articulated bodies with an arbitrary number of links, including loops.
In the following, $\mcA_a$ will denote an assembly of $M_a$ rigid bodies forming an articulated rigid body with $P_a$ links, so that $M = \sum_{a=1}^N M_a$ and $P = \sum_{a=1}^N P_a$ correspond to the total number of rigid bodies and links respectively. We also define $\mcL_p$, the set of links attached to the rigid body $p$.
Each inelastic link connects two bodies by a hinge joint, see Fig. \ref{fig:sketch_joint}.
The links, while inextensible, can rotate.
Therefore, the $n^{\tex{th}}$ link, connecting bodies $p$ and $q$, enforces the \new{vectorial} constraint
\eqn{
  \label{eq:constraint}
  \bg_n(\bx) = \bq_p + \bR(\btheta_p) \bDl_{np} - \bq_q - \bR(\btheta_q) \bDl_{nq} = \bs{0},
}
where $\bDl_{np}$ is the vector from the body $p$ to the hinge $n$ in the body frame of reference.
This vector is then rotated to the laboratory frame of reference by the
rotation matrix $\bR(\btheta_p)$, see Eq.\ \eqref{eq:rot_mat}.
Passive links are time independent in the body frame of reference, however, we will allow for
time dependencies, i.e.\ $\bDl_{np}=\bDl_{np}(t)$, to simulate active links.

For systems with $M$ rigid bodies and $P$ links it is convenient to write all the constraints in one equation
\eqn{
   \label{eq:constraintCompact}
   \bg = \bP \bq + \bz(\btheta) = \bs{0}.
}
The vector $\bz_n(\btheta) = \bR(\btheta_p) \bDl_{np} - \bR(\btheta_q) \bDl_{nq}$ carries the orientation dependent
terms while the sparse matrix $\bP$, of size $3P \times 3M$, encodes the bodies connectivity.
If bonds are neither formed nor broken during a simulation $\bP$ remains constant.

The nonlinearity of \eqref{eq:constraintCompact} with respect to the particles' orientations poses a challenge to integrate the equations of motion \cite{Featherstone1987,Schoeller2021}.
However, as the constraints are holonomic their time derivatives are linear in the velocities of the rigid bodies
\eqn{
  \label{eq:dg}
  \dot{\bg}_n &= \bu_p + \bomega_p \times (\bR(\btheta_p) \bDl_{np}) - \corchete{\bu_q + \bomega_q \times (\bR(\btheta_q) \bDl_{nq})} - \bB_n(t)=\bs{0}, 
}
where $\bB_n(t)= \bR(\btheta_q) \dot{\Delta \bs{l}}_{nq}(t) - \bR(\btheta_p) \dot{\Delta \bs{l}}_{np}(t)$ is zero for passive links.
As working with linear constraints is much simpler we will use \eqref{eq:dg} to derive the equations of motion.
Once more we introduce a more compact notation
\eqn{
  \label{eq:dg_compact}
  \dot{\bg} = \bC \bU  - \bB(t) = \bs{0},
}
where the matrix $\bC = (\partial \dot{\bg} / \partial \bU)$ is the constraints' Jacobian.
Each constraint exerts a force and torque on the rigid bodies which in the case of passive links, $\bC \bU=\bB(t)=0$, generate no work \cite{Goldstein2002}.
This condition is enough to determine the structure of the constraint generalized forces,
$\bF^C = \bC^T \bphi$, where $\bphi$ is a Lagrange multiplier \cite{Delmotte2015b}.
It is easy to verify that the generated power, $dW = \bU \cdot \bF^C = \bphi \cdot \bB(t)$, indeed vanishes for passive links and that the  constraint forces do not exert any net force or torque on the whole articulated body, see \ref{sec:convervation_momentum}.
For articulated bodies the balance of force and torque reads
\eqn{
  \label{eq:balance}
  \bs{\mc{K}}^T \blambda - \bC^T \bphi = \bF.
}
Equations \eqref{eq:Stokes}-\eqref{eq:no-slip-continuum} together with \eqref{eq:dg_compact}-\eqref{eq:balance} form
a linear system that describe the dynamics of articulated rigid bodies immersed in a Stokes flow.
At this point, it is worth mentioning that, without loss of generality, the constraint equation \eqref{eq:constraint} can apply to a single spatial component instead of being vectorial.
In addition, it can involve the position and/or orientation of a single body instead of two, without changing the formalism presented above.

It is interesting to use \eqref{eq:U_NF} to write the linear system with constraints as
\eqn{
  \label{eq:linear_system_general}
\left[\begin{array}{cc}
\new{-}\bN \bC^T& \bI  \\
 \bzero & \bC 
\end{array} \right]
\left[\begin{array}{c}
\bphi \\
\bU 
\end{array} \right] =
\left[\begin{array}{c}
\bN \bF \\
\bB
\end{array} \right],
}
where $\bI$ is a $6M \times 6M$ identity matrix.
Any  method to solve the Stokes problem, and therefore to apply the mobility matrix $\bN$, can be used with \eqref{eq:linear_system_general}.

It can be enlightening to write the formal solution of the linear system \eqref{eq:linear_system_general} to gain some physical intuition on the equations.
First, in the absence of links (\new{$\bC=\bzero$}) we can write the velocity of the rigid bodies as $\what{\bU} = \bN \bF$.
Then, when the bodies are linked, the velocity can be written as 
\eqn{
  \label{eq:U}
  \bU &= \what{\bU} + \bN \bC^T \bphi, \nonumber \\
  &= \corchete{\bI - \bN \bC^T \bG \bC} \what{\bU} + \bN \bC^T \bG \bB(t),
}
where $\bG = \pare{\bC \bN \bC^T}^{-1}$ is the \emph{constraint resistance matrix}.
Both lines of \eqref{eq:U} are enlightening.
The first line shows that the unconstrained velocity is rectified by the flow generated by the constraint forces.
It also shows that the linear system becomes a mixed mobility-resistance problem, as the constraint forces, $\bphi$, are unknowns of the linear system.
The second line shows that for passive links, i.e.\ $\bB(t) = 0$, the constrained velocity can be found by projecting $\what{\bU}$ to the space of admissible velocities.
Note that the \new{matrix} $\bQ = {\bI - \bN \bC^T \bG \bC}$ is a projection operator, i.e.\ $\bQ^2 = \bQ$.

\subsection{Rigid Multiblob Model as a Stokes solver}
\label{sec:rigid_multiblob}

Solving the mobility-resistance problem applying an iterative solver to \eqref{eq:linear_system_general} could be quite expensive as, in general, computing
the action of the mobility $\bN$ requires solving a linear system of its own.
We seek a formulation that avoids nested linear systems to reduce the computational cost. 
For that reason we will solve the Stokes problem and the constraint equations
simultaneously.
To solve the hydrodynamic problem we rely on the rigid multiblob method \cite{Usabiaga2016}. 
We discretize the surface of each rigid body by a set of markers or \emph{blobs} of finite radius $a$ and position $\br_i$.
\new{Fig. \ref{fig:sketch_joint}b shows an example of discretization of a bacterium with a spherical head and a helical flagellum.}
Each blob is subject to a force, $\blambda_i$, that enforces the rigid motion of the body.
Thus, the integrals in the balance of force and torque \eqref{eq:balance} become sums over the blobs
\eqn{
\label{eq:balanceF}
\sum_{i\in \mcB_p} \blambda_i -\sum_{n \in \mcL_p} \bphi_n  &= \bbf_p, \\
\label{eq:balanceT}
\sum_{i \in \mcB_p} (\br_i - \bq_p) \times \blambda_i - \sum_{n \in \mcL_p} (\bDl_{np} - \bq_p) \times \bphi_n &= \btau_p,
}
where the second sum runs over the links \new{$\mcL_p$} attached to the rigid body $p$ and $\bphi_n$ is the constraint  Lagrange multiplier acting on link $n$.

The no-slip condition, as in collocation methods \cite{Pozrikidis1992}, is evaluated at each blob
\eqn{
\label{eq:no-slip}
\bv(\br_i) =  \sum_j \bM_{ij} \blambda_j &= \bu_p + \bomega_p \times (\br_i - \bq_p) \;\; \mbox{for all } i \in \mcB_p, 
}
where the mobility matrix $\bM$ mediates the hydrodynamic interactions.
The term $\bM_{ij}$ couples the force acting on the blob $j$ to the velocity of blob $i$.
In a first-kind boundary integral formulation the mobility would be simply the Green's function of
the Stokes equation \cite{Pozrikidis1992, Bao2018}.
The rigid multiblob method uses instead a regularization of the Green's function, the so-called Rotne-Prager approximation \cite{Rotne1969},
which can be written as the \new{double} integral of the Green's function, $\bG(\br)$, over the blobs surface \cite{Wajnryb2013}
\eqn{
  \label{eq:RPY}
  \bM_{ij} = \bM(\br_i, \br_j)  &=  \fr{1}{(4\pi a^2)^2}  \int \delta(|\br' - \br_i|-a) \bG(\br', \br'') \delta(|\br'' - \br_j|-a) \dd^3 r'' \dd^3 r',
}
where $\delta(\br)$ is the Dirac's delta function and $a$ is the blob radius.
The advantage of \eqref{eq:RPY} over the non-regularized Green's function is that the RPY mobility is always positive definite, even when blobs overlap.
This property makes the numerical method quite robust and easy to implement.
There are analytical expressions to calculate the RPY mobility in some geometries such as unbounded spaces \cite{Rotne1969, Wajnryb2013}
or above an infinite no-slip wall \cite{Swan2007}.
Moreover, there are fast methods to compute the product $\bM\blambda$ in quasilinear time in those domains \cite{Liang2013, Yan2018a}
as well as in periodic domains \cite{Fiore2017,Maxey2001}.

The whole linear system to solve the  \new{constrained} mobility problem is found by combining \eqref{eq:balanceF}-\eqref{eq:no-slip},
\eqn{
\label{eq:linear_system}
\underbrace{
\left[\begin{array}{ccc}
\bM & -\bK & \bzero \\
-\bK^T & \bzero & \bC^T \\
\bzero & \bC & \bzero 
\end{array} \right]}_{\bA}
\underbrace{
\left[\begin{array}{c}
\blambda \\
\bU \\
\bphi
\end{array} \right] }_{\by}=
\underbrace{
\left[\begin{array}{c}
\bu_s \\
-\bF \\
\bB
\end{array} \right]}_{\bb}.
}
In the linear system the matrix $\bK$ is a discretization of the operator $\bs{\mc{K}}$, so it transforms the rigid body velocities to blobs velocities.
Note that in the right hand (RHS) side we have included a slip velocity, $\bu_s$, on the blobs.
To solve \eqref{eq:linear_system} with iterative methods, such as GMRES, it is necessary that the implementation of the equations
can handle nonzero slips.
Besides, $\bu_s$ can be used to model active slip on the bodies, e.g.\ to model a squirmer \cite{Blake1971a} or a phoretic particle \cite{Brosseau2019, Brosseau2021}.
The solution of \eqref{eq:linear_system} is identical to \eqref{eq:U} except that the unconstrained body velocities now contain a slip-induced term
\eqn{
\what{\bU} = 
\bN\bF \new{-} \bN\bK^T\bM^{-1}\bu_S
}
where $\bN = \left(\bK^T\bM^{-1}\bK\right)^{-1}$ is the body mobility matrix given by the rigid multiblob model.

\subsection{Constraints with single blobs}
\label{sec:single_blobs}

For some specific applications it is convenient to represent rigid bodies as single blobs.
For example, an inextensible filament can be modeled by discretizing the centerline with  blobs \cite{Yamamoto1993,Delmotte2015b,Schoeller2021} or
a sheet can be formed by blobs linked to their nearest neighbors \cite{Silmore2021}.

At this level of approximation the application of the mobility matrix, $\bN$, simply requires computing hydrodynamic interactions between blobs, including the coupling with the rotational degrees of freedom. 
This can be achieved with analytical approximations of the generalized RPY tensor \cite{Wajnryb2013} or by using grid-based Stokes solvers with the Force Coupling Method \cite{Lomholt2003} or similar techniques.
Therefore, the computational cost is much lower and the velocities of the rigid articulated  bodies  can be found by solving \eqref{eq:linear_system_general} with an iterative method such as GMRES.
We will use this approach in Section \ref{sec:convergence_results} to model filaments and spherical deformable shells.


\section{Preconditionner and iterative solver convergence}
\label{sec:precond_conv}
\refone{In this section we first present the iterative method used to solve the constrained mobility problem \eqref{eq:linear_system}.
  In order to ensure and speed-up the convergence of the iterative solver, we use a new block diagonal preconditioner
  that generalizes  one introduced for freely suspended, unconstrained, rigid bodies \cite{Usabiaga2016}.
  Then  we evaluate its effectiveness on linear articulated bodies, such as filaments, and closed ones with loops, such as shells.
  While these two examples cover a broad range of interesting applications, this section focuses on the solver performances rather than the physical phenomena at stake. }
\subsection{Block diagonal preconditionner}
\label{sec:precond}
In the following, we use the preconditioned GMRES method to solve the constrained mobility problem \eqref{eq:linear_system} iteratively.
Due to the long-ranged nature of hydrodynamic interactions at low Reynolds numbers, the condition number of the matrix $\bA$ in  Eq.\ \eqref{eq:linear_system} increases with the blob number and volume fraction, which slows down the convergence of iterative solvers for large and/or concentrated suspensions.

As shown in our previous work \cite{Usabiaga2016}, the convergence \refthree{on the unconstrained mobility problem} can be improved with a very effective, yet simple, left preconditioner $\wtil{\bA}^{-1}$ considering non-interacting bodies. \refthree{Applying $\wtil{\bA}^{-1}$ to the unconstrained mobility problem with the rigid multiblob method amounts to solving the approximate problem \cite{Usabiaga2016}
   \eqn{
\left[\begin{array}{ccc}
\wtil{\bM} & -\bK  \\
-\bK^T & \bzero \\
\end{array} \right]
\left[\begin{array}{c}
\blambda \\
\bU 
\end{array} \right] =
\left[\begin{array}{c}
\bu_s \\
-\bF
\end{array} \right]
}
where $\wtil{\bM}^{(pq)} =\delta_{pq}\bM^{(pp)}$ is nonzero only for blobs belonging  to the same body $p$. The corresponding body mobility matrix $\wtil{\bN}^{(pq)}$ is block diagonal,  where each block  refers to a single body neglecting all hydrodynamic interactions with other bodies 
\eqn{
\wtil{\bN}^{(pq)} =\delta_{pq}\left[\left(\bK^{(p)}\right)^T\bM^{(pp)}\bK^{(p)}\right]^{-1}.
}}
\refthree{Doing so, we found that, for freely suspended bodies,} the number of iterations  becomes independent of the number of bodies, and slightly increases  with the volume fraction \cite{Usabiaga2016}.

In this work we  \refone{apply this approach}  to the constrained system \eqref{eq:linear_system}, which amounts to solving the approximate problem
    \eqn{
    \label{eq:precond}
\left[\begin{array}{ccc}
\wtil{\bM} & -\bK & \bzero \\
-\bK^T & \bzero & \bC^T \\
\bzero & \bC & \bzero 
\end{array} \right]
\left[\begin{array}{c}
\blambda \\
\bU \\
\bphi
\end{array} \right] =
\left[\begin{array}{c}
\bu_s \\
-\bF \\
\bB
\end{array} \right].
}

As a result, the approximate problem can be solved  for each articulated body separately, and in parallel, since the resulting approximate constraint resistance matrix is block diagonal for each assembly 
\eqn{
\wtil{\bG}^{(\mcA_a \mcA_b)} =\delta_{ab}\left[\bC^{(\mcA_a)}\wtil{\bN}^{(\mcA_a\mcA_a)}\left(\bC^{(\mcA_a)}\right)^T\right]^{-1}.
\label{eq:block_diag_const_res}
}
If the matrix $\bC$ does not have full row rank, $\wtil{\bG}$ is evaluated with the pseudo-inverse.

\subsection{Convergence results}
\label{sec:convergence_results}
In the following we evaluate  the effectiveness of the preconditioned iterative solver on linear articulated bodies, such as filaments, and closed ones with loops and branches such as shells.
\refone{We show that, just like unconstrained bodies, the number of iterations does not grow with the system size and therefore that our scheme is effective to perform many body simulations of articulated bodies.}

\subsubsection{Settling filaments}

We consider an array of initially straight, inextensible, and deformable filaments \refone{(i.e.\ with zero bending stiffness)} sedimenting in the $(x,z)-$plane in free space, \new{with gravity pointing in the $-z$ direction}.
\refone{The dynamics of sedimenting fibers has been thoroughly investigated at the individual \cite{Marchetti2018, Delmotte2015b} and collective level \cite{Herzhaft1999,Butler2002,Guazzelli2011,Schoeller2021,Nazockdast2017}. Here we use this example exclusively to investigate the performance of our iterative solver with linear chains of constraints and refer the reader to the cited articles for more details on the modelling aspects and the physics of the system.}\\
Each filament $\mcA_a$ is oriented along the $x$-axis and  discretized with $M_a = 15$ spheres with hydrodynamic radius $a = 1\, \si{\mu m}$. 
Discretizing  filaments with spheres is very common in the literature and it has proven to accurately and efficiently capture their dynamics  at the individual and collective level \cite{Ross1997,Gauger2006,Lindstrom2007,Delmotte2015b,Marchetti2018,Schoeller2021,Zuk2021}.
The inextensibility constraint of a filament states that the derivative of the centerline position $\bq$ with respect to the arclength $s$ must correspond to a unit vector $\hat{\bt}$ tangent to the centerline:
\eqn{
\frac{\partial \bq}{\partial s} = \hat{\bt}.
\label{eq:inext_cont}
}
After discretizing the filament with $M_a$ bodies and $M_a-1$ links, Eq.\ \eqref{eq:inext_cont} becomes
\eqn{
\bq_{p+1} - \bq_{p}  -  \frac{l_c}{2}\left(\hat{\bt}_p     + \hat{\bt}_{p+1}\right)  = \bzero, \, \forall p \in \mcA_a,
\label{eq:inext_cont_disc}
}
where $\hat{\bt}_p $ is the unit orientation vector of body $p$ along the filament centerline, and $l_c$ is the length of the link connecting the centers of body $p$ and $p+1$. 
Using the general notation introduced in Section \ref{sec:kin_const},   Eq.\ \eqref{eq:inext_cont_disc} writes
\eqn{
 \bq_{p+1} + \bR(\btheta_{p+1})\cdot \bDl_{n,p+1} - \bq_p  - \bR(\btheta_{p})\cdot \bDl_{n,p} = \bzero
 \label{eq:inext_const_filament}
}
where the link vectors in the body frame are  $\bDl_{n,p+1} = - \bDl_{n,p} =  -l_c \be_x/2$ 
for a straight filament initially aligned along the $x$-axis.

\begin{figure}
 \centering
\includegraphics[width=0.95 \columnwidth]{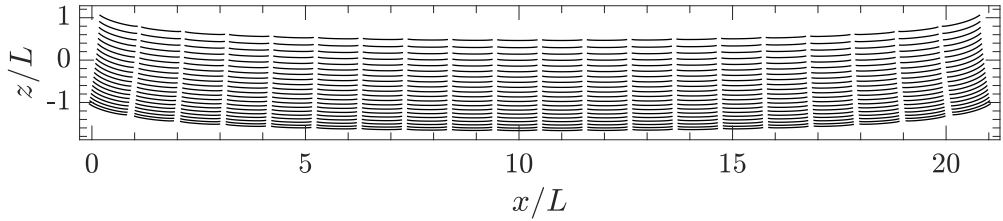}
\caption{
  Snapshot of a sedimenting  array of $20\times 20 = 400$ filaments, discretized with $M_a = 15$ spheres, 
  \refone{bended by the hydrodynamic drag and the flow disturbances induced by their neighbours \cite{Schoeller2021}.
    Snapshot} at dimensionless time $t\times U_z/L =0.09$, where $U_z = -3.26\, \si{\mu m/s}$ is
  the velocity of an isolated straight filament subject to the gravitational force.
}
  \label{fig:snapshot_filaments}
\end{figure}

In the following we set the link length between two bodies $l_c = 2.5a$ so that the filament length and aspect ratio are $L = (M_a-1)\times l_c + 2a = 37a$ and  $L/2a =18.5$ respectively.  
The total number of filaments in the array is $N = N_x\times N_z$ where $N_x$ and $N_z$ are the number of filaments along each direction. 
The spacing between filaments along the $x$ and $z$ direction is  $4a$, so that the local area fraction is approximately  $\phi \approx 0.32$.
Each body is subject to a gravitational force along the $z$-axis with magnitude  $mg=2.5 \cdot 10^4\, \si{pN}$.
The solvent viscosity is $\eta = 10^3$ Pa$\cdot$s.
Figure \ref{fig:snapshot_filaments} shows a snapshot of the sedimenting array \refone{at the area fraction considered in this test.}

The spheres making up the filaments are either modeled as single blobs or as a rigid icosahedron made of 12 blobs  distributed on the sphere surface. The number of filaments is varied from $N=1$ up to $N=20\times20=400$, which corresponds to a maximum system size of $52,800$ for single blobs and $268,800$ for the multiblob model.

\begin{figure}
 \centering
\includegraphics[width=0.45 \columnwidth]{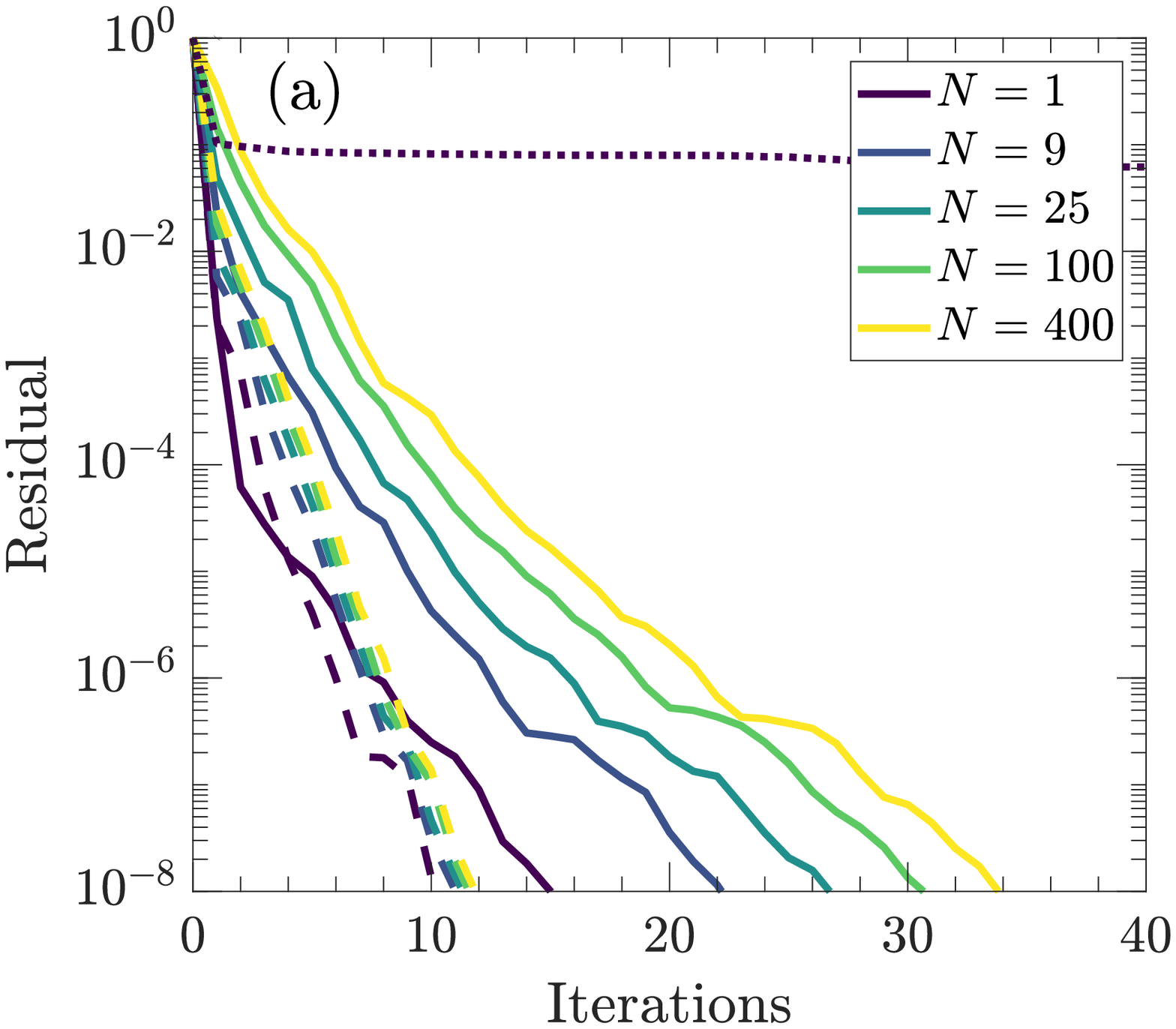}
\includegraphics[width=0.45 \columnwidth]{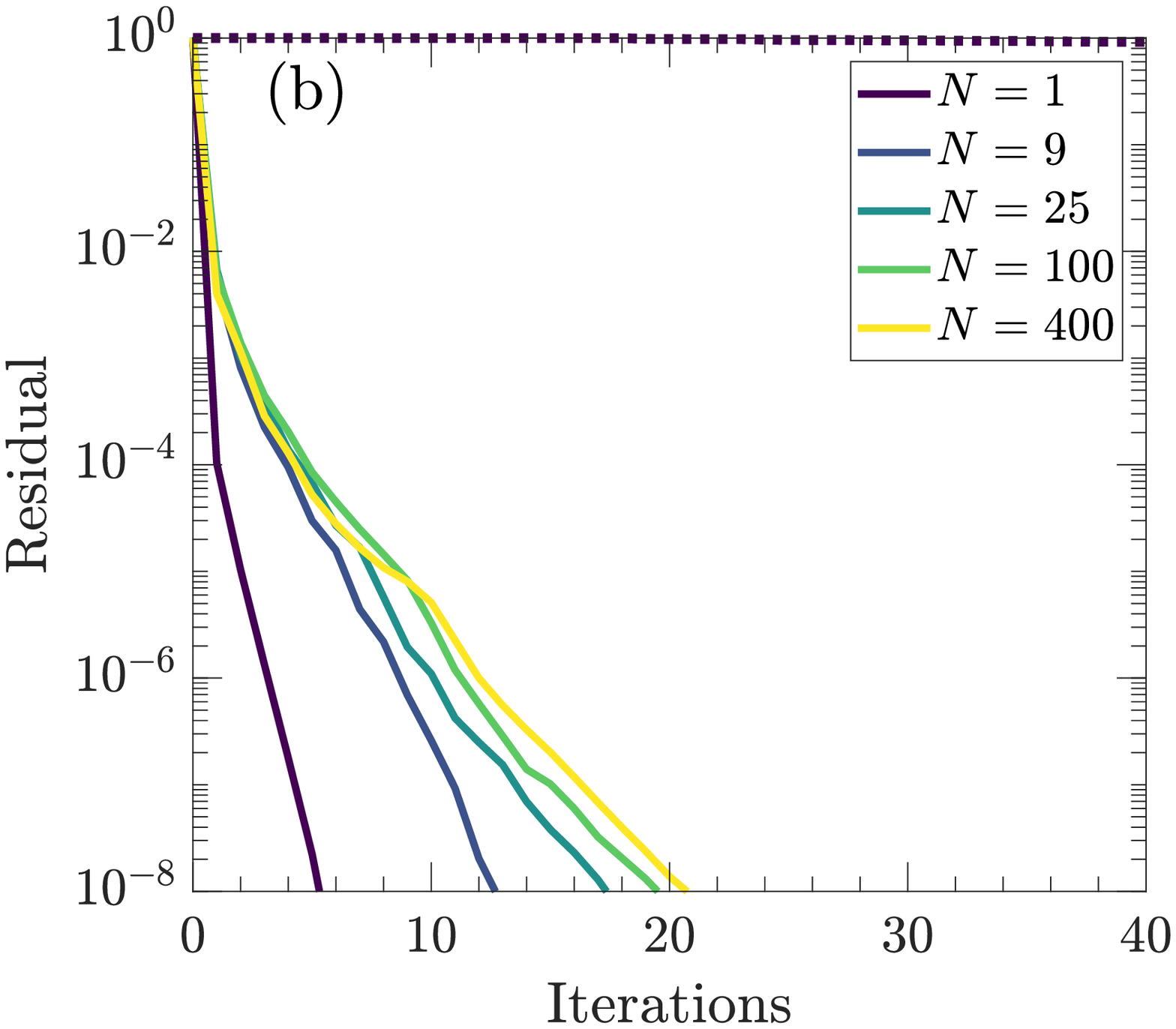}\\
\includegraphics[width=0.45 \columnwidth]{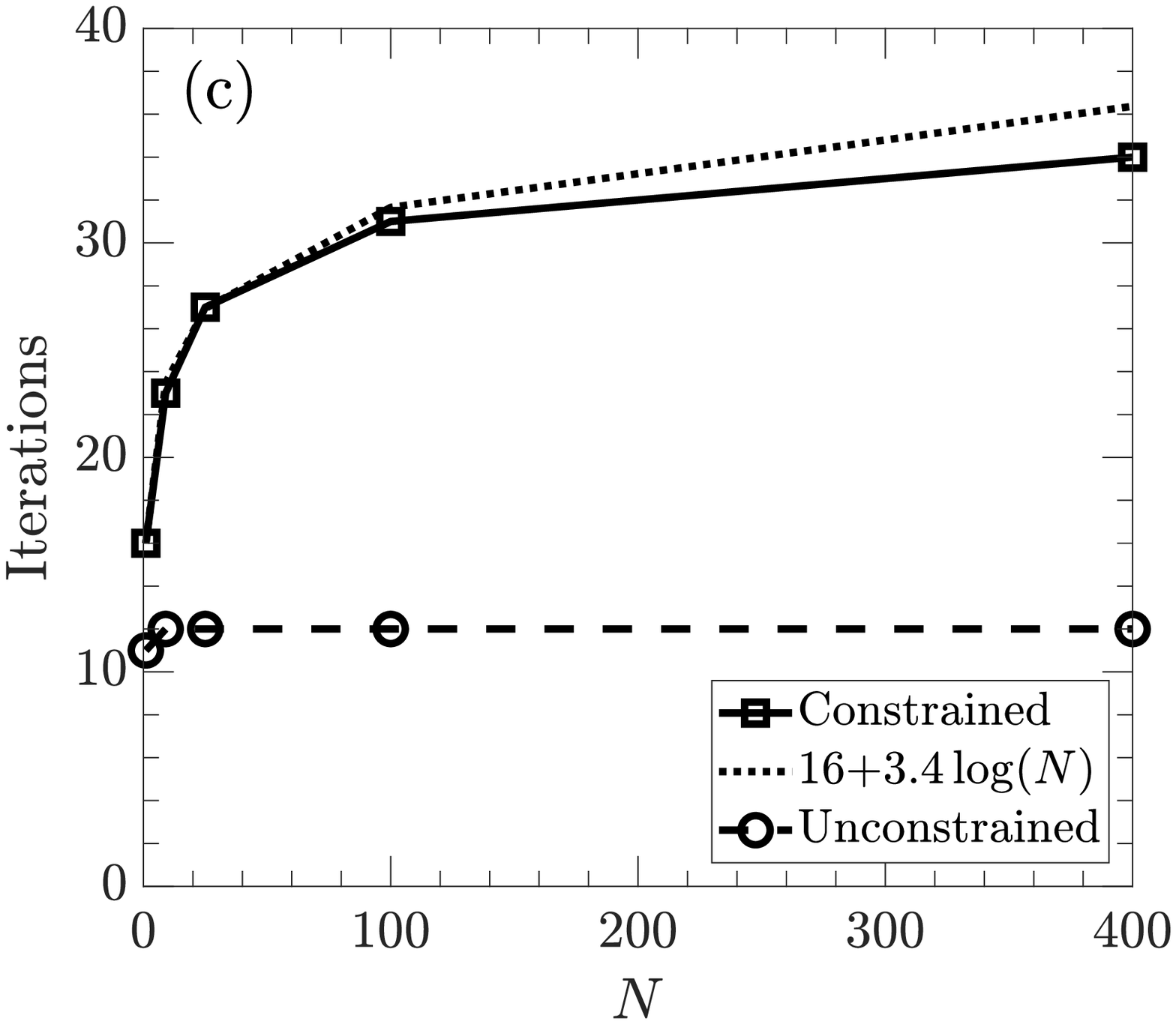}
\includegraphics[width=0.45 \columnwidth]{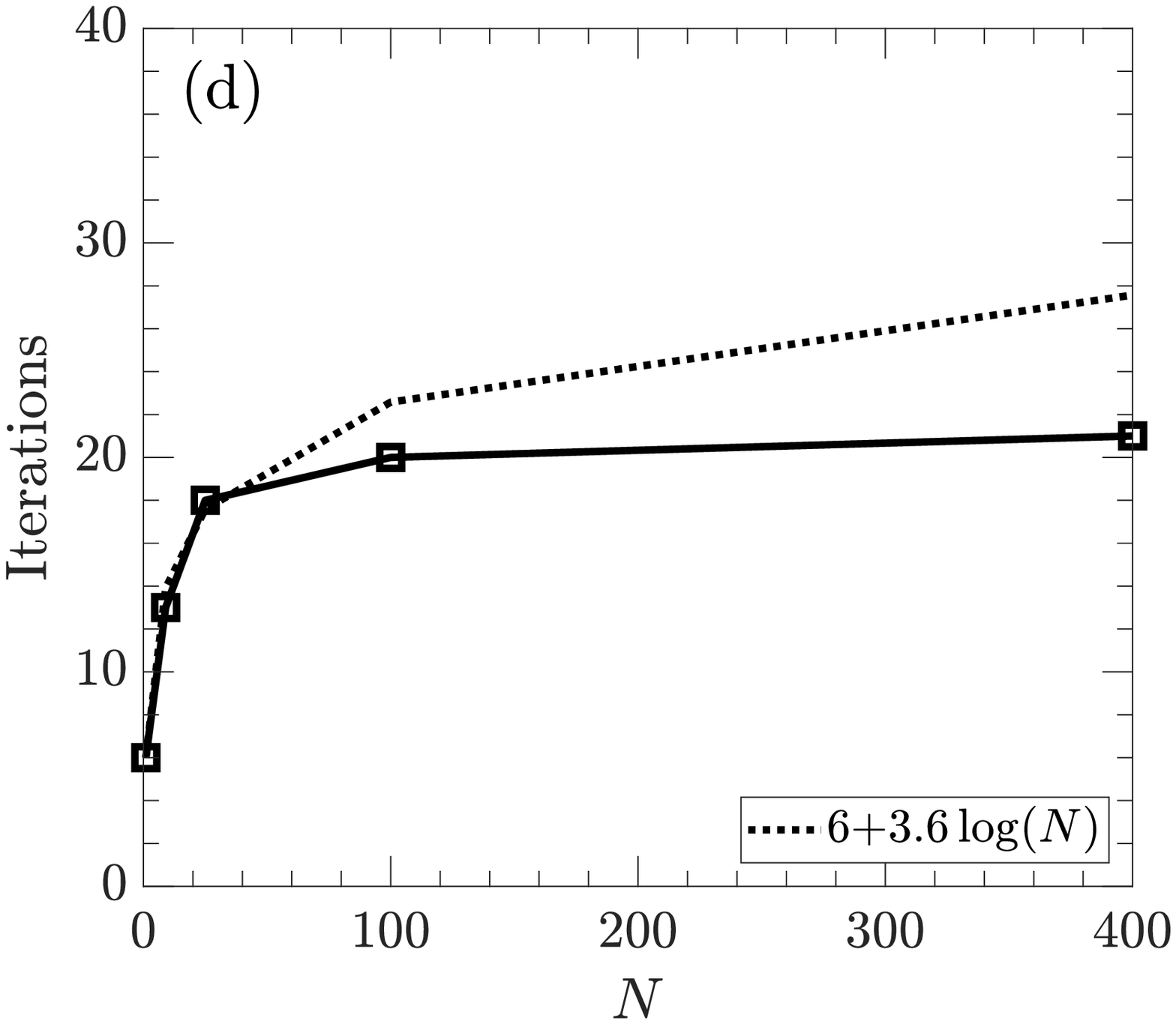}
\caption{Linear solver convergence for different number of sedimenting filaments, $N$, discretized with $M_a = 15$ icosahedra (a,c) or blobs (b,d) each. The dashed lines in panels (a,c) correspond to an identical configuration with unconstrained icosahedra. The dotted line in panel (a,b) correspond to the convergence of GMRES without preconditioner with $N=1$ filament. Panels (c,d) show the number of iterations required to reach a tolerance  $\epsilon = 10^{-8}$. The dotted lines represent a logarithmic fit of the first three data points.}
  \label{fig:convergence_filaments}
\end{figure}

Before evaluating the performance of our preconditioner, we stress that even with only one filament, the GMRES solver  converges extremely slowly  in the absence of preconditioner,  regardless of the discretization  (Fig.\ \ref{fig:convergence_filaments}a,b dotted line). 
We have found that the typical number of iterations required to reach an acceptable tolerance is approximately equal to the system size which leads to a cubic scaling similar to  direct methods, hence the need for a good, scalable preconditioner.

Fig.\ \ref{fig:convergence_filaments}  shows the convergence of  the preconditioned  iterative solver for the two levels of discretization.
First we examine the effect of including the inextensibility constraint \eqref{eq:inext_const_filament} between spheres  on the solver convergence.
In the absence of kinematic constraints the spheres are free to move and the preconditioned iterative solver for the multiblob model  converges within 10 iterations independently of the number of filaments $N$ (Fig.\ \ref{fig:convergence_filaments}a), which is expected from our previous work \cite{Usabiaga2016}. When the inextensibility constraints are included, the convergence slows down with  $N$. However, as shown in Fig.\ \ref{fig:convergence_filaments}c,  the number of iterations to reach a tight tolerance  ($\epsilon = 10^{-8}$) scales sub-logarithmically with the particle number.  
For the single-blob model (Fig.\ \ref{fig:convergence_filaments}b,d), the convergence rate is faster and quickly becomes independent of the number of filaments.  
The independence on the particle number seems to be a generic feature of the preconditioner since a similar plateau is observed for arrays of deformable shells, as shown in the next section.

Altogether, these results demonstrate the ability and effectiveness of the preconditioner to handle constrained systems. 
When all the degrees of freedom of the bodies are constrained, i.e.\ \new{$N_c = N_{dof} = 6M$ where $N_c$ is the number of independent constraints}, Eq.\  \eqref{eq:linear_system} becomes a resistance problem, where one needs to invert the whole mobility matrix $\bN$ to obtain the forces corresponding to the prescribed particle velocities. It has been shown that preconditioned  iterative solvers for resistance problems scale, at best, sublinearly as $M^\alpha$, where $0<\alpha<1$  \cite{Ichiki2002,Usabiaga2016}.
Our constrained problem is a \textit{mixed} resistance-mobility problem since the ratio of \new{independent} constraints over the number of degrees of freedom of the system is smaller than unity: for a filament discretized with $M_a \in \llbracket 2,\infty \llbracket$ spheres the ratio is in the range \new{$N_c/N_{dof} = 3(M_a-1)/6M_a = 0.25 - 0.5$}.  In order to find the constraint forces, one needs to invert the  constraint mobility matrix $\bC\bN\bC^T$, which is smaller than $\bN$.
It is therefore interesting to note that when the number of constraints is smaller than the number of degrees of freedom our preconditioned iterative solver becomes independent of the particle number ($\alpha=0$), which is a significant improvement over sublinear scaling. 

Note that in practical applications, a looser tolerance is usually employed ($\epsilon  = 10^{-3}-10^{-4}$), thus lowering the number of iterations to 6 - 14  with the multiblob model and only \new{3-4}  with single blobs, which drastically reduces the computational cost. 
\new{Note also that, in dynamic simulations, the iterative solver uses the solution from the previous time-step as a first guess. By doing so, the number of iterations is further reduced: in the example shown in Fig.\ \ref{fig:snapshot_filaments} only 1-2 iterations were necessary to reach $\epsilon  = 10^{-4}$ with the single blob discretrization.}

\subsubsection{Deformable shells}

\begin{figure}
   \includegraphics[width=0.95 \columnwidth]{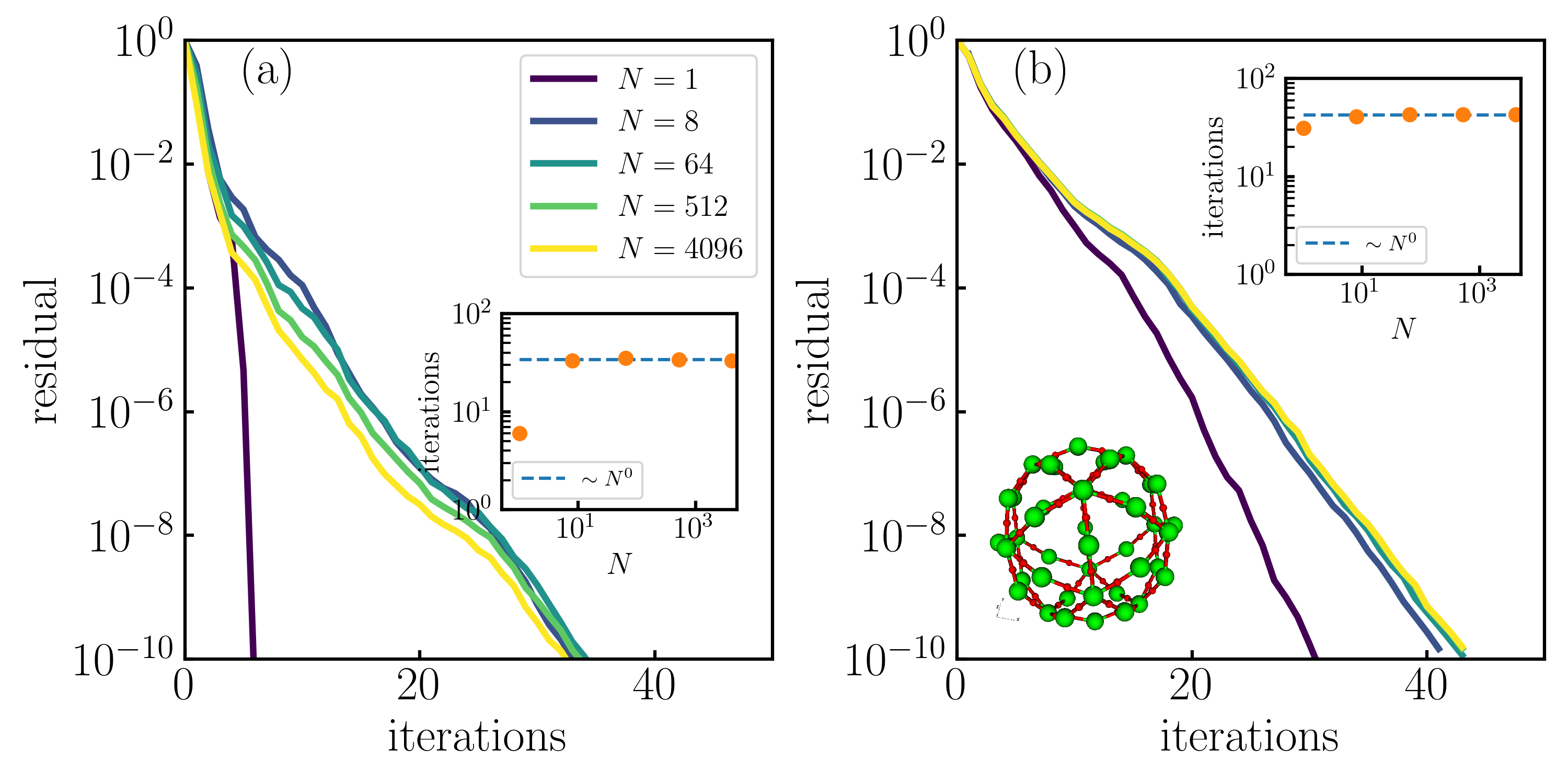}
  \caption{Linear solver convergence for different number of shells, $N$, under (a) constant force or (b) random forces and torques. Insets show the number of iterations required to reach a tolerance $\epsilon = 10^{-10}$. Shells are discretized with $M_a=42$ blobs and $P_a=60$ links, see inset in (b).}
  \label{fig:convergence_shells}
\end{figure}

We further test the robustness of our preconditioner on deformable shells \refone{(i.e.\ with zero bending stiffness)} where, contrary to filaments, the links  form loops.
\refone{
  Such complex constraints arrangements appear in any discretization of inextensible surfaces, like membranes,
  but here we just focus on the solver performance and not on the physical aspects of the model.}
Each shell is discretized with $M_a = 42$ blobs with minimal spacing $l_c \approx 0.25$ times the shell diameter.
The blobs are linked to their first neighbors.
Most blobs have only two first neighbors but some have five
so that the total number of links per shell is $P_a = 60$, see inset of Fig. \ref{fig:convergence_shells}b.
For an articulated shell, the ratio of the number of constraints   over the number of degrees of freedom is \new{$N_c / N_{dof} = 3\times 60/(42\times6) = 0.71$} which is closer to a  resistance problem (for which this ratio would be exactly one) compared to the filament case above. 
We consider a cubic lattice of $N = 1 - 4096$ shells two diameters apart which corresponds to a local volume fraction of $\phi \approx 0.065$. 
The maximum system size of the linear system, reached for $N=4096$, is $1,769,472$.
Figure \ref{fig:convergence_shells} shows the convergence of the preconditioned iterative solver with two types of forcing on the blobs: a constant force (e.g.\ due to gravity, Fig.\ \ref{fig:convergence_shells}a) and random forces and torques (e.g.\ due to thermal fluctuations, Fig.\ \ref{fig:convergence_shells}b).
As in the filament simulations, the number of iterations to reach a given tolerance
does not depend on the number of shells regardless of the forcing type (see insets), even though the ratio constraints/degrees of freedom is closer to unity.
Owing to its complexity, the random forcing case  requires more iterations to converge than the sedimenting case (e.g.\ 16 vs.\ 8 iterations to reach a residual of $10^{-4}$).
Altogether, these results confirm the robustness  of the preconditioner and scalability of the iterative solver for the constrained mobility problem.

 \section{Body reconstruction and time-integration}
 \label{sec:time-integrator}
As explained in Section \ref{sec:kin_const}, the kinematic constraints on the velocities are obtained by taking the time derivative of the position constraints. Thanks to this transformation, the constrained problem is reduced to a single linear system for the body velocities that can be efficiently solved with our  preconditioned GMRES solver (cf.\ Section \ref{sec:precond_conv}). However, satisfying the velocity constraints does not guarantee that the position constraints will be obeyed with the same accuracy. Indeed, when integrating Eqs.\ \eqref{eq:dq_dt}-\eqref{eq:dtheta_dt} for the particle positions and orientations, errors of order $O(\Delta t^{q+1})$, where $q$ is the order of convergence of the discrete time integrator, will accumulate at each time step. The error in the position constraints may therefore become uncontrolled as time advances.
To circumvent that problem, we propose \refone{to track the articulated bodies with a new} reconstruction method and a correction procedure that satisfy the position constraint up to an arbitrary precision while preserving the  $O(\Delta t^{q})$  global error of the time integration scheme. 

\subsection{Reconstruction method for general articulated bodies}

\subsubsection{The  robot-arm model \refone{for linear chains}}
A well-known approach to prevent error accumulation in linear chains, such as filaments,  is the  ``robot-arm" model. Reconstructing a robot-arm only requires tracking an arbitrary point of the articulated bodies (e.g.\ the first body) and the orientation of each link composing the arm. In the case of a single robot-arm with tracking point $\bq_1$, the position of body $p>1$ is given by simply adding the link vectors along the chain
\begin{eqnarray}
\bq_p & = & \bq_1 + \sum_{q=1}^{p-1} \corchete{\bR(\btheta_{q})\cdot \bDl_{q+1,q} -\bR(\btheta_{q+1})\cdot \bDl_{q,q+1}},
\label{eq:robot-arm}
\end{eqnarray}
where $\bDl_{q,q+1}$ (respectively $\bDl_{q+1,q}$) is the vector connecting body $q$ (resp.\ $q+1$) to the hinge between $q$ and $q+1$.
Therefore, the robot-arm parametrization ensures that the body positions satisfy the position  constraint exactly regardless of the integration error on the tracking point $\bq_1$ and the orientations $\btheta = \{\btheta_p\}_{p=1}^M$.
However, the reconstruction \eqref{eq:robot-arm} does not work for general articulated bodies with branches and/or loops.

\subsubsection{General reconstruction method}
\refone{Here we present our extension of the robot-arm model to arbitrary constraint arrangements.}
Since articulated bodies $\{\mcA_a\}_{ a = 1}^{N}$ can have  arbitrary connections, a natural choice for the tracking point is their center of mass (COM) $\bq_{COM} = \{\bq^a_{COM}\}_{a = 1}^{N}$.
Therefore, instead of integrating the position for all the rigid bodies \eqref{eq:dq_dt}, the equations to be integrated in time are 
\begin{eqnarray}
\frac{d\bq_{COM}}{dt} &= &\bu_{COM}, \label{eq:eq-motion-reconst-COM}\\
\frac{d\btheta_p}{dt} &= &\fr{1}{2} \corchete{0, \bomega_p} \bullet \btheta_p, \,\, p=1,..,M,
\label{eq:eq-motion-reconst-orient}
\end{eqnarray}
where $\bu_{COM} =\{\bu^a_{COM}\}_{a = 1}^{N} $ with
\eqn{
\bu^a_{COM}  = \frac{1}{M_a}\sum_{p\in \mc{A}_a}\bu_p,
\label{eq:ucom}
}
the mean velocity of articulated body $\mcA_a$. 

A simple way  to generalize the robot-arm model for articulated bodies including loops and branches is to solve the connectivity problem given by the matrix form of the  constraints \eqref{eq:constraintCompact} for each articulated body separately
\eqn{
\bP^a\bq^a = -\bz^a(\btheta,t), \, a = 1,..,N,
\label{eq:Pq_eq_z}
}
where $\bP^a$, $\bq^a = \left\{\bq_p\right\}_{p\in\mcA_a}$ and $\bz^a$ are  the $3P_a \times 3M_a$ connectivity matrix, $3M_a\times 1$ body positions and $3P_a\times 1$ relative orientations associated to articulated body $a$.
Since each constraint involves pairs of bodies, and because the number of \new{links} per articulated body always satisfies $P_a\geq (M_a-1)$, the rank of $\bP^a$ is $r= 3(M_a-1)$ regardless of the constraint arrangement.
Therefore Eq.\ \eqref{eq:Pq_eq_z} has infinitely many solutions.  The nullspace of $\bP^a$ is spanned by  $3M_a-r = 3$ basis vectors. Each of these basis vector contains a single position repeated $M_a$ times, which corresponds to a shift of the COM of the articulated body. In order to remain in the frame attached to the articulated body, we remove all the nullspace contributions by solving \eqref{eq:Pq_eq_z} with the pseudo-inverse of $\bP^a$
\eqn{
\wtil{\bq}^a = -\left(\bP^a\right)^{+}\bz^a.
\label{eq:pseudo-inv-sol}
}
One can check that the resulting solution has indeed zero COM: $\wtil{\bq}^a_{COM} =  \frac{1}{M_a}\sum_{p\in \mc{A}_a}\wtil{\bq}_p = \bzero$.
The articulated body is then reconstructed by adding the position of the COM obtained after integrating \eqref{eq:eq-motion-reconst-COM}
\eqn{
\bq_p = \wtil{\bq}_p + \bq^a_{COM}, \, \forall p \in \mcA_a, \, a = 1,..,N.
\label{eq:reconst-up-COM}
}
\new{Since arbitrary translations of the COM are removed with the pseudo-inverse in \eqref{eq:pseudo-inv-sol}, the reconstruction \eqref{eq:reconst-up-COM} is unique regardless of the constraint arrangement: there is only one way to construct an assembly with a given set of links and a given reference position. Therefore, for linear chains, our general  reconstruction method is strictly identical to the robot-arm model, regardless of the choice of the tracking point.}

\subsection{Correction step for discrete time-integrators}
If \eqref{eq:eq-motion-reconst-COM}-\eqref{eq:eq-motion-reconst-orient} are integrated exactly in time, then $\wtil{\bq}^a$ satisfies the position constraints and so do the reconstructed positions in \eqref{eq:reconst-up-COM}.
However, in practical simulations, \eqref{eq:eq-motion-reconst-COM}-\eqref{eq:eq-motion-reconst-orient} are integrated with a discrete time integrator.
After integrating over a time-step, the resulting  $O(\Delta t^{q+1})$ error on the body orientations $\btheta_p$, is transmitted to the RHS of Eq.\ \eqref{eq:Pq_eq_z}.
If the articulated bodies do not have any loops, e.g.\ branched filaments, the solution  \eqref{eq:pseudo-inv-sol} exactly obeys the constraints
independently of the time-stepping errors in the RHS.
However, when the constraints form loops, the solution  $\wtil{\bq}$ might violate the position constraints.

\begin{figure}
\includegraphics[width=0.95 \columnwidth]{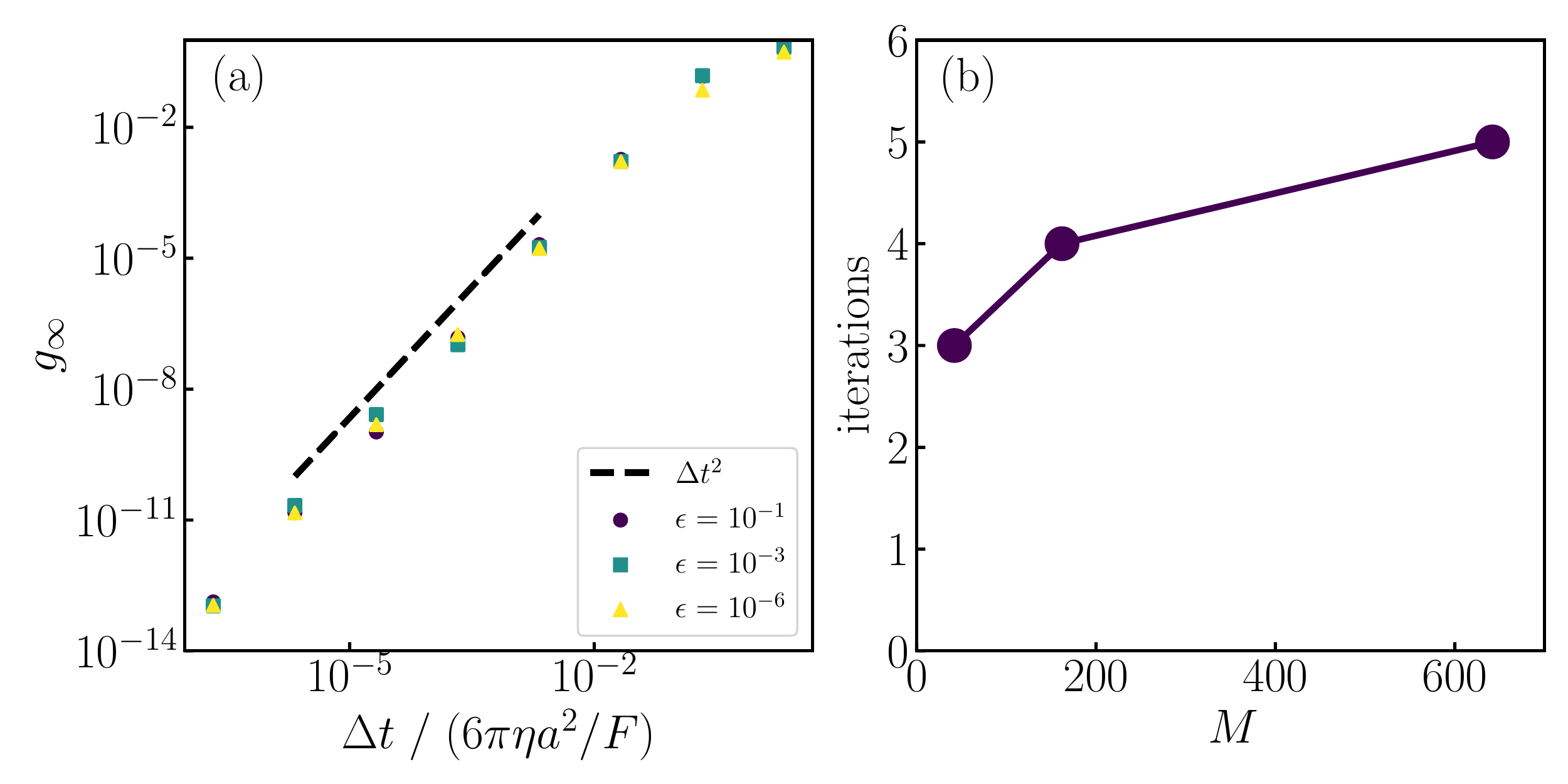}
\caption{Constraint errors and nonlinear convergence for a shell formed by blobs under random forces and torques.
    (a) Infinite norm of the constraint error before the correction step versus the integration time step size $\dt$  for a shell formed by $M = 42$ blobs and $P = 60$ links.
  Using the preconditioner the velocity obeys the constraint for any tolerance $\epsilon$ and therefore the error decreases as the $O(\dt^2)$ local truncation error of Explicit Euler.
  (b) Number of iterations in the nonlinear solver to converge to a tolerance $\delta = 10^{-10}$ versus the number of blobs $M$. The number of blobs in the shell is $M=42$, $162$ and $642$ and the number of links $P=60$, $240$ and $963$ respectively.}
  \label{fig:convergence_nonlinear}
\end{figure}

We quantify the constraint violation due to a finite time-step by simulating a deformable shell made of $M=42$ blobs and $P=60$ links. Each blob constituting the shell is subject to random forces and torques. The equations of motion \eqref{eq:eq-motion-reconst-COM}-\eqref{eq:eq-motion-reconst-orient} are integrated with an  Explicit Euler  scheme over one time-step.
Fig.\ \ref{fig:convergence_nonlinear}a shows the convergence of the position constraint error, here defined as the infinity norm of the constraint vector $\|\bg\|_{\infty}$, with $\Delta t$ for several linear solver tolerances $\epsilon = 10^{-6} - 10^{-1}$.
As the preconditioner guarantees that the velocities obey the constraints for any $\epsilon$, the constraint violations show second order convergence,
as expected for an Explicit Euler integrator.
Even though the constraint error remains low and decays as $\Delta t^2$, it is preferable to prevent its accumulation and to decorrelate  it from the time discretization.\\

To prevent any constraint violation at the end of the time step we correct the body positions and orientations with small increments,
$\bdq = \{\bdq_p\}_{p=1}^{M}$ and $\bdth = \{\bdth_p\}_{p=1}^{M}$ respectively. These increments are  solutions to the minimization problem
\eqn{
  \label{eq:min_probl_a}
  \left(\breve{\bdq}, \breve{\bdth}\right) &=  \mbox{arg}\min \| \bg(\bdq,\bdth)\|_2^2 \\
  \label{eq:min_probl_b}
  &\mbox{s.t. } \|\bdth_p\|_2^2 - 1 = 0,\, \forall p, 
}    
where 
\eqn{
  \label{eq:residual}
  \bg_{n}(\bdq,\bdth) = \bq_p + \bdq_p  + \bR(\bdth_{p}\bullet\btheta_p) \bDl_{np} - \bq_q - \bdq_q -  \bR(\bdth_{q}\bullet\btheta_q) \bDl_{nq},
}
is the $n^{th}$ constraint linking bodies $p$ and $q$.
The second equation, \eqref{eq:min_probl_b}, constrains the quaternion increments to have unit norm to properly represent
rotations\footnote{We could have written the minimization problem with rotation vectors $\bgamma_p$ so that
  $\bdth_p = \btheta_{\bgamma_p} = [\cos(\gamma_p/2), \sin(\gamma_p/2)\hat{\bgamma}_p]$, the quaternion obtained from the vector
  $\bgamma_p = \gamma_p \hat{\bgamma}_p$, automatically satisfies the unit norm constraint.
  However, we found that approach to be less effective in terms of computational cost due to the cumbersome expressions of the
  Jacobian matrix $\bJ = \partial\bg/\partial\bdx$.}.

We solve \eqref{eq:min_probl_a}-\eqref{eq:min_probl_b} using the least squares algorithm implemented in the \emph{Scipy} library \cite{Virtanen2020}.
We have observed that the correction is proportional to the local truncation error of the time-integration scheme,
thus the nonlinear solver does not affect the integrator global convergence rate.
It is possible to add inequalities to the minimization problem to ensure the corrections does not exceed the local truncation error of the time-integration scheme
\cite{Voglis2004}.
We use the exact Jacobian of the residual to speedup the convergence, see \ref{sec:Jacobian};
in this case the number of residual and Jacobian evaluations is equal to the number of iterations plus one.
Since the Jacobian is very sparse its use does not increase the memory requirements of the algorithm appreciably. 
The number of iterations to attain a tight tolerance, e.g.\ $\delta = 10^{-10}$, is almost independent of the size of the articulated body.
Figure \ref{fig:convergence_nonlinear}b shows that the least square solver converges in a small number of iterations even for large articulated bodies with many constraints.

\subsection{Reconstruction and correction with Explicit Euler}
Our reconstruction and correction methods combines effectively with any time integration scheme. For the sake of simplicity, we outline the algorithm for the Explicit Euler integrator.
An algorithm with an Explicit Midpoint scheme is provided in \ref{app:midpoint}.

\begin{itemize}
    \item Initialize positions $\bq(0)$, $\bq_{COM}(0)$ and orientations $\btheta(0)$ 
    \item Precompute the pseudo-inverse of the connectivity matrix $\bP^+$.
    \item Time loop: for $k = 0,..,N_{it}-1$
    \begin{enumerate}
        \item Solve the linear system \eqref{eq:linear_system} at $t_k = k\Delta t$ to obtain the body velocities $\bU^{(k)}$.

        \item Compute the translational velocity of the COMs  $\bu_{COM}^{(k)}$ using \eqref{eq:ucom}.
 
        \item Update the COM   and the orientations of each assembly
        \begin{eqnarray}
        \bq_{COM}^{\star} &=& \bq_{COM}^{(k)} + \Delta t \bu^{(k)}_{COM},\\
        \btheta^{\star} &=& \btheta_{\bomega^{(k)}\Delta t}\bullet\btheta^{(k)}.
        \end{eqnarray}
  
        \item Reconstruct the articulated bodies using the pseudo-inverse \eqref{eq:pseudo-inv-sol}, where  $\bz^a = \bz^a(\btheta^\star,t_{k+1})$   and then apply Eq.\  \eqref{eq:reconst-up-COM} to obtain the body positions $\bq^{\star}$ attached to the updated centers of mass.
        \item Evaluate the updated constraint vector $\bg^\star = \bg(\bq^\star,\btheta^\star,t_{k+1})$ and  check constraint violation:\\ 
        if $\|\bg^\star\|^2_2 < \delta$ then 
        \eqn{
        \bq^{n+1} \leftarrow \bq^{\star},\\
        \btheta^{n+1} \leftarrow \btheta^{\star},
        }
        else
        \begin{itemize}
        \item Solve the minimization problem \eqref{eq:min_probl_a}-\eqref{eq:min_probl_b} with the nonlinear solver to find the increments $(\breve{\bdq}, \breve{\bdth})$.
        \item Correct the body positions and orientations
          \begin{eqnarray}
            \bq^{n+1} &=& \bq^{\star} + \breve{\bdq},\\
            \btheta^{n+1} &=& \breve{\bdth}\bullet\btheta^{\star}.
          \end{eqnarray}
        \end{itemize}
    \end{enumerate}
\end{itemize}
We note that the correction step is not necessary for open chains since our reconstruction method satisfies the constraints regardless of the time discretization errors.

\subsection{Convergence in time}
\label{sec:time_convergence}
Before running dynamic simulations, we confirm that our method preserves the global error of the time integration scheme  to which it is combined, here Explicit Euler. 
   For long enough simulations, we expect the linear solver tolerance $\epsilon$ to shift the value of the global error, without affecting its convergence rate. Indeed, since the body velocities are obtained by solving \eqref{eq:linear_system} iteratively with a tolerance $\epsilon$, the position and orientation updates acquire an error $O(\epsilon \dt)$ at each Explicit Euler time step. After a simulation time $t_f$, the accumulated error due to the iterative solver is therefore $O(\epsilon t_f)$, with a prefactor that depends on the system configuration. \refone{Below we use two examples, a swimming bacterium and the sedimentation of a pair of shells, to check that our reconstruction and correction steps do not affect the convergence of the discrete time-integrator.}

\begin{figure}
 \centering
\includegraphics[width=0.45 \columnwidth]{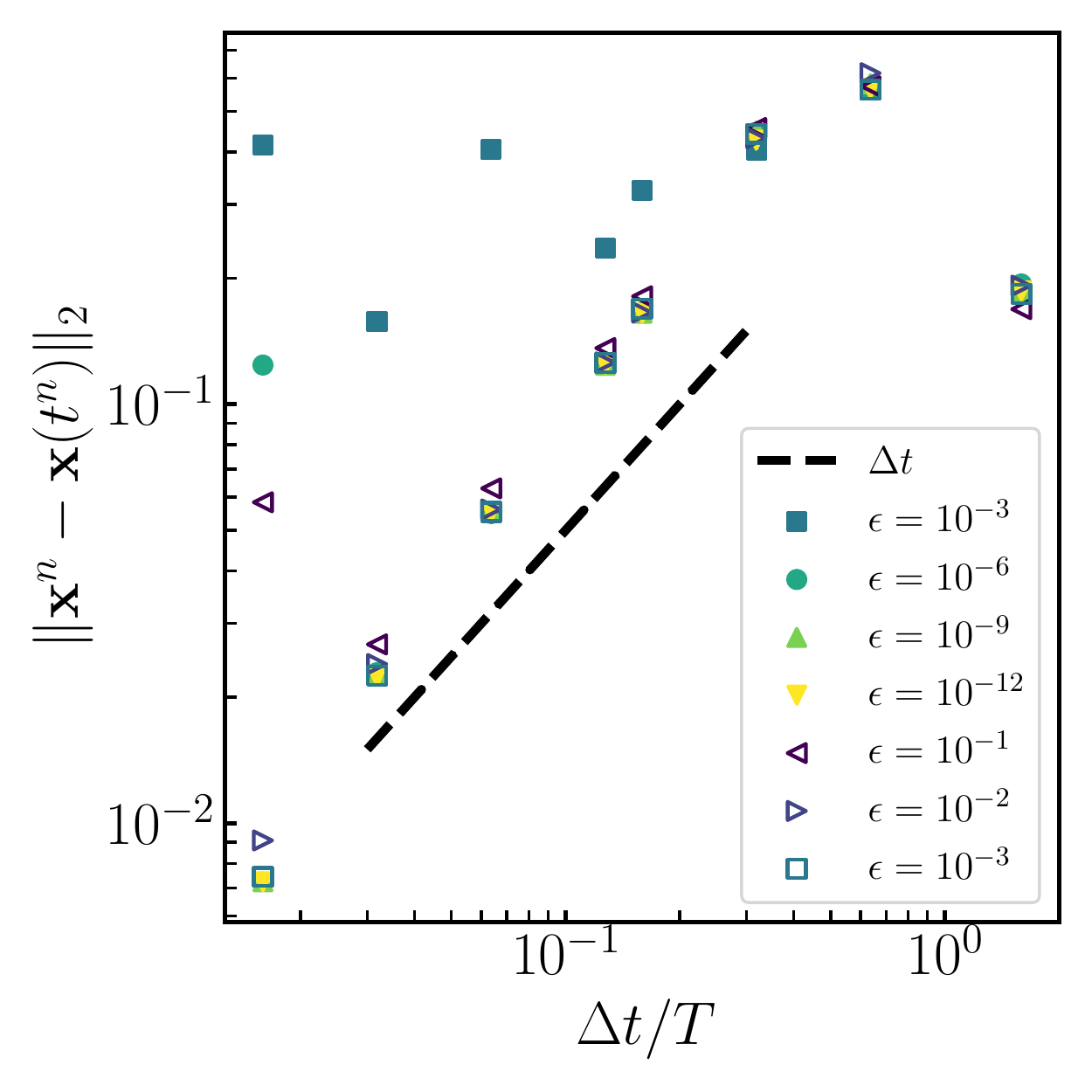}
\includegraphics[width=0.46 \columnwidth]{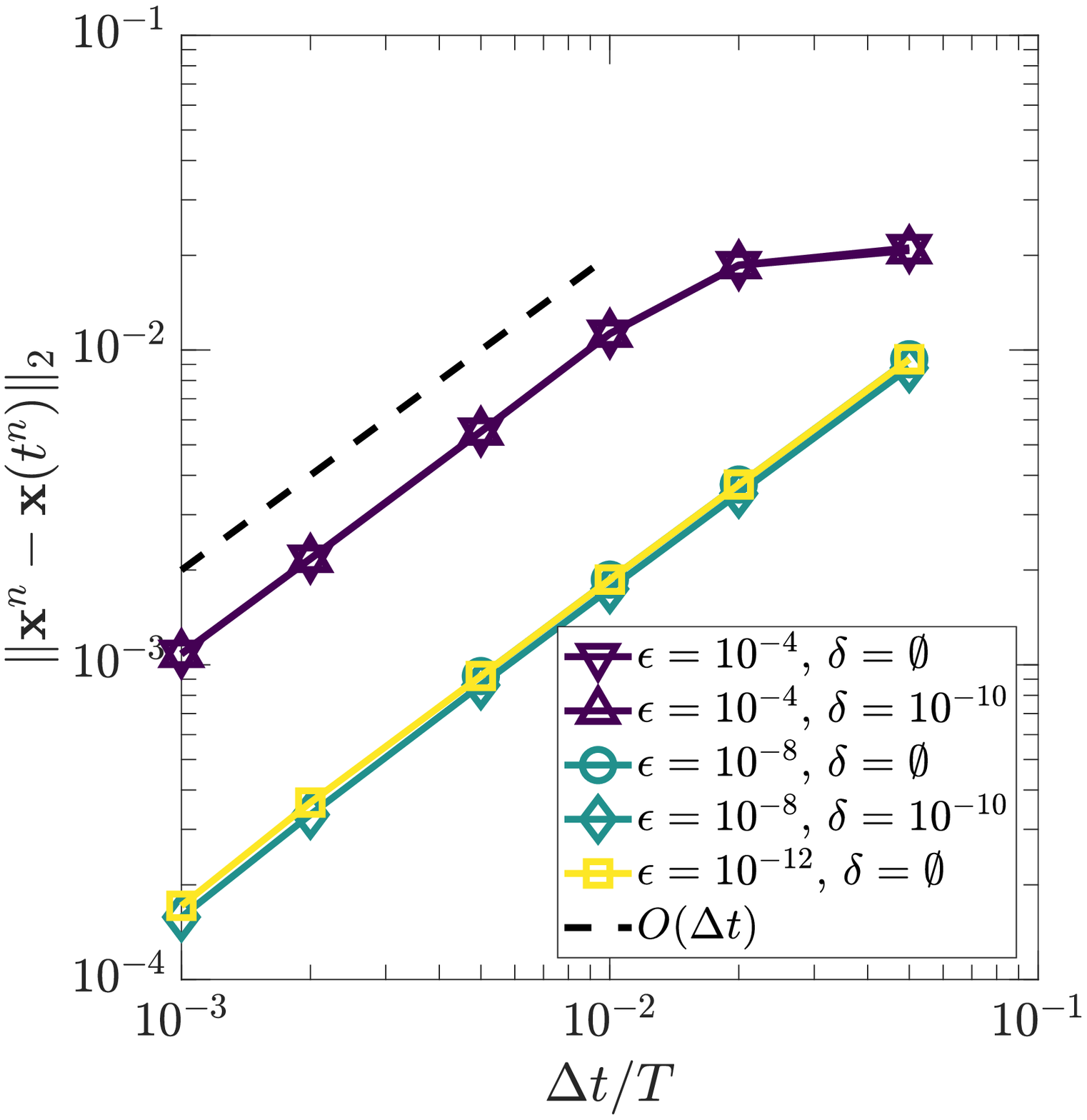}
\caption{Time convergence of the Explicit Euler scheme quantified by the distance to a  reference solution at final time $t^n=t_f$. (a) Results for a swimming bacterium. Full symbols: same tolerance for linear and nonlinear solver ($\epsilon = \delta$). Open symbols: tight tolerance for nonlinear solver ($\delta = 10^{-12}$).
  First order convergence is observed when the constraint violations are small ($\delta \leq 10^{-9}$). 
  (b) Convergence for two sedimenting shells with ($\delta = 10^{-10}$) and without ($\delta = \emptyset$) correction.
  First order convergence is always observed.
  }
  \label{fig:convergence_time_step}
\end{figure}

First we consider  the motion of a swimming bacterium.
The bacterium is formed by a spherical body of radius $R=1\,\si{\mu m}$ and a single helical flagellum of length $L=10\,\si{\mu m}$, see Fig. \ref{fig:bacteria_fields}. 
We discretize the spherical body with $162$ blobs of radius $a=0.131\,\si{\mu m}$ and the flagellum with only $38$ blobs.
The effective thickness of the flagellum is \refone{$h \approx a/2$}.
We incorporate two links between the spherical body and the flagellum,
\new{
\eqn{ 
  \bg_n(\bx) &= \bq_{\text{body}} + \bR(\btheta_{\text{body}}) \bDl_{n\,\text{body}} - \corchete{\bq_{\text{flagellum}} + \bR(\btheta_{\text{flagellum}}) \bDl_{n\,\text{flagellum}}} = \bzero, \\
  \bDl_{1\,\text{body}} &= (0, 0, R+a), \;\;\;
  \bDl_{1\,\text{flagellum}} = (0, 0, -a), \\
  \bDl_{2\,\text{body}} &= (0, 0, 2R+a), \;\;\;
  \bDl_{2\,\text{flagellum}} = (0, 0, R-a),
}}
the first keeps the flagellum attached at the body surface while the second,
\new{at an arbitrary distance from the first but aligned along the same direction,}
fixes the axis of rotation.
Then we apply equal but opposite torques ($\tau=0.4644\, \si{pN}\cdot \si{\mu m}$) to the body and the flagellum parallel to the axis of rotation,
as the flagellum rotates the bacterium swims forward.      
We simulate the swimming bacterium for about $16$ rotations periods $T$ which moves the bacterium a distance $d\approx 2.5 R$.
The  matrix $\bC$ corresponding to the constraints in Eq.\ \eqref{eq:dg_compact} does not have full row rank (rank 5 instead of 6)
\new{as the links are not linearly independent.}
The resulting linear system is not invertible as such and the iterative solver alone cannot converge.
As indicated in Section  \ref{sec:precond}, the preconditioner computes the block-diagonal constraint resistance matrix using a pseudo-inverse when $\bC$ is not full rank to achieve convergence.
However, the resulting solution is sensitive to small constraint violations: a constraint violation as small as $\|\bg\|=10^{-6}$ at time iteration $t_k$ can lead to large discrepancies between the preconditioned residual $\wtil{r} = \|\wtil{\bA}^{-1}(\bA\bx-\bb)\| \approx 10^{-12} $ and the actual residual $r = \|\bA\bx-\bb\| \approx 10^{-3}$ at time iteration $t_{k+1}$.
In figure Fig.\ \ref{fig:convergence_time_step}a we show the time convergence with respect to a reference simulation with tight tolerances $\epsilon=\delta=10^{-12}$ and a small time step $\dt / T = 8\cdot 10^{-3}$.
Due to the error accumulation the expected $O(\dt)$ convergence with Explicit Euler is not observed for a loose nonlinear solver tolerance $\delta>10^{-9}$ (filled symbols).
However, as soon as the constraint tolerance is tight enough $\delta\le 10^{-9}$ (open symbols), the scheme converges as expected, regardless of the linear solver tolerance. 
  Identical tests using an explicit midpoint scheme show similar results: the $O(\dt^2)$ convergence is observed as long as $\delta$ is small enough (not shown). 
  Using a tight tolerance $\delta$ for the constraint violation is possible because the cost of the nonlinear solver is negligible compared to solving the linear system, and only a few iterations, typically less than 5, are required per articulated body, see Fig.\ \ref{fig:convergence_nonlinear}b.
When $\delta \leq 10^{-9}$, the effect of the linear solver tolerance is visible as the global error slightly  decreases with $\epsilon$ (open symbols).

  Our second example is the sedimentation of $N=2$ deformable shells with  radius $R$  made of $M_a=42$ spheres and $P_a = 60$ links each. Spheres are discretized as single blobs. In this case the  links  are all independent and the corresponding matrix $\bC$ has full row rank. The shells are initially placed at positions $(\pm 1.5,0,\pm 1.5)R$ and sediment for $t_f = 5T$ where $T = R/U_0$ where $U_0$ is the settling speed at $t=0$. Fig.\ \ref{fig:convergence_time_step}b  reports the maximum $L_2$ error of the final body positions and orientations compared to a reference solution obtained with a very small time-step size, $\Delta t/T  = 10^{-4}$, and tight tolerances, $\epsilon = \delta = 10^{-12}$. Since the constraints are independent, the preconditioned GMRES solver is not sensitive to constraint violations.
  For this specific configuration the shift due to the $O(\epsilon t_f)$ error from the linear solver is clearly visible  between $\epsilon = 10^{-4}$ and $\epsilon = 10^{-8}$. 
  We also observe that the constraint violations are so small that the effect of the correction scheme is only visible below a tight tolerance: when $\epsilon \leq 10^{-8}$  the correction step slightly decreases the global error and even allows to obtain more accurate results than  uncorrected positions with $\epsilon  = 10^{-12}$.

 Altogether, these results show that our correction procedure preserves the accuracy of the time-integration scheme regardless of the constraint arrangement.

\section{Simulations}
In this Section we simulate two different kinds of active systems to further illustrate the flexibility and versatility of our method. We describe, \refone{for the first time,} the various locomotion modes of a model swimmer inspired by the diatom chain \textit{Bacillaria Paxillifer}. Then we investigate the swimming speed of a model bacterium, \refone{compare our results with theoretical models,} and finally exploit the scalability of our solver to simulate large suspensions of bacteria near a no-slip surface.

\subsection{Swimming motion of a model diatom chain}
\begin{figure}
 \centering
 \includegraphics[width=0.99 \columnwidth]{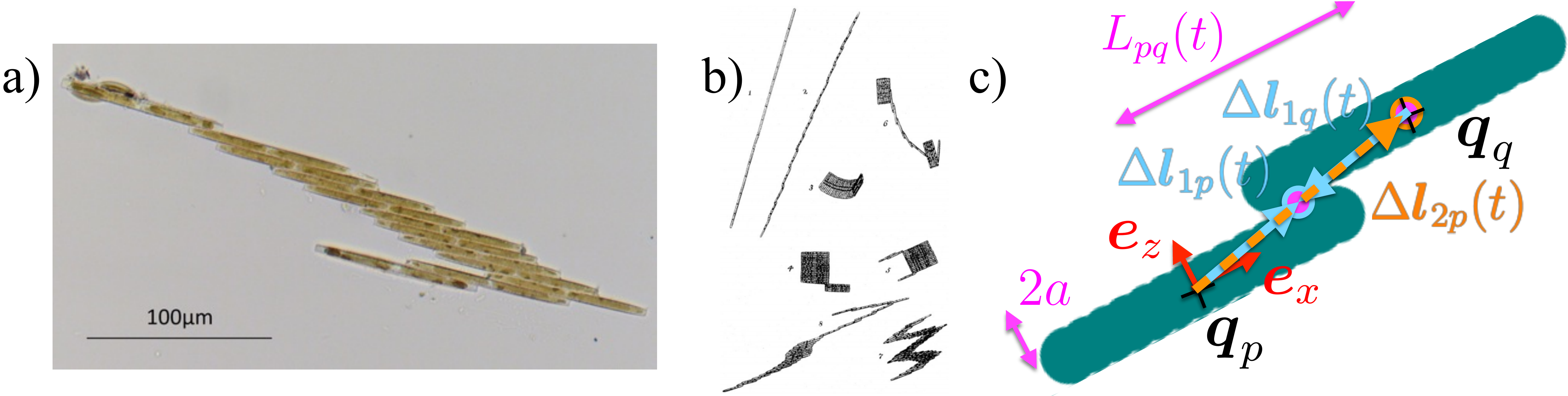}
  \caption{a) Microscope view of a colony made of $M=12$ rod-like cells \cite{dskeet}. b) Drawing of the various configurations observed by O.F Müller, who discovered the species \textit{Bacillaria Paxillifer} in 1783 \cite{Muller1782}.  c) Schematic of the \refone{two links} enforcing the kinematic sliding constraint between two adjacent rods $p$ and $q$. \refone{Cyan arrows: connections to link 1  \eqref{eq:const_bacillaria_b}. Orange arrow: connection to link 2  \eqref{eq:const_bacillaria_c}. Magenta  circles: location of the links.} The basis vectors represent the local frame attached to body $p$. }
  \label{fig:const_bacillaria}
\end{figure}

\textit{Bacillaria Paxillifer} is a diatom species found in a wide variety of natural environments such as  marine and freshwaters \cite{Jahn2007}. Cells are rectangular in shape with typical length $L \approx 70 \mu$m, height and width $H \approx W \approx 10 \mu$m, and live in stacked colonies (see Fig.\ \ref{fig:const_bacillaria}a). Colonies of diatoms are phototactic and motile. Members, with their long axes parallel to one another, slide against their neighbors in a coordinated fashion, allowing the structure to expand or contract in many different ways (see Fig.\ \ref{fig:const_bacillaria}a,b). 
\refone{So far, the diversity of deformation sequences, their purpose and their mechanical origin remain a mystery. The literature on these subjects is very scarce. In this work, we carry out the first numerical simulation of such microorganisms and show that hydrodynamic interactions between sliding cells lead to various, nontrivial, locomotion modes.}\\
Our model swimmer inspired by \textit{Bacillaria Paxillifer} is made of $M = 16$ rigid rods of aspect ratio $L/H = 6.3$ connected together by kinematic constraints in the $(x,z)-$plane. Each rod is discretized with 14 blobs of size $a = H/2$  separated by a distance  $0.81a$, which was shown to provide highly accurate predictions of the rod mobility coefficients \cite{Bringley2008,Usabiaga2016}. The length of the stacked colony is therefore $L_c = M \times H = 16 \times 2a$.
The kinematic constraints  prescribe a time-dependent parallel sliding motion \refone{and a constant normal distance} between pairs of rods.
\refone{To enforce both of these constraints, two links are incorporated between each pair of adjacent rods $p$ and $q$:
\eqn{
\bg_n(\bx) &= \bq_p + \bR(\btheta_p) \bDl_{np} - \left[\bq_q + \bR(\btheta_q) \bDl_{nq}\right] = \bs{0}  \label{eq:const_bacillaria},\\
\bDl_{1p} & = 0.5 L_{pq}(t) \be_x + a\be_z,\;\;\;\; \bDl_{1q}= -\bDl_{1p},\label{eq:const_bacillaria_b}\\
  \bDl_{2p} & = L_{pq}(t) \be_x + 2a\be_z,\;\;\;\;  \bDl_{2q}= \bzero,\label{eq:const_bacillaria_c}\\
\mbox{with }   &  L_{pq}(t) = A \sin(2\pi ft + \phi_n),  
}
}
where $A = 1.8L$ is the sliding amplitude, and $\phi_n = (n-1) \Delta \phi, n=1,\dots,M-1$, is the local phase,  $\Delta \phi$ being the phase shift between two adjacent pairs of sliding rods. \refone{The location of the second link in eq.\ \eqref{eq:const_bacillaria_c} is arbitrary as long as it lies on the axis connecting $\bq_p$ and $\bq_q$ and does not coincide with the first link.} This relative sliding motion can be seen as a deformation wave in the frame oriented with the colony: $z(x_n + ct) \sim \sin(k(x_n + ct))$ with wave number $k = 2\pi/\lambda = (M-1)\Delta \phi/L_c$, speed  $c = 2\pi f/k$ along the $-x$ direction, and discrete horizontal positions $x_n = (n-1)2a, n=1,..,M$. However, unlike typical travelling waves, the amplitude of the deformation wave depends on the the wavelength $\lambda$ because of the relative sliding motion: the shorter the wavelength, the larger the deformation.

\begin{figure}
  \centering
  \includegraphics[width=1 \columnwidth]{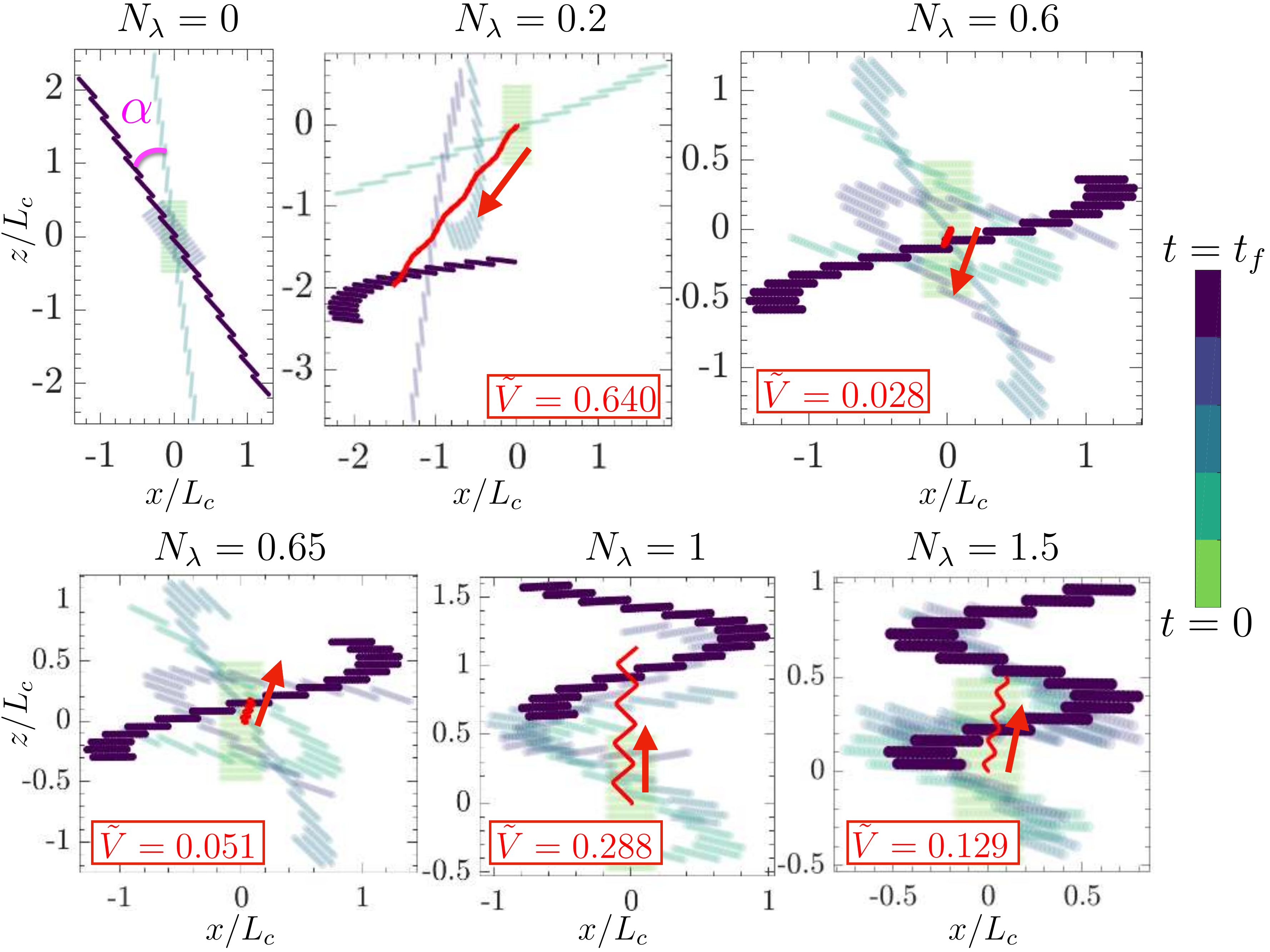}
  \caption{Trajectory and  conformation of the colony for six values of the number of wavelengths along the colony $N_\lambda = L_c/\lambda = 0 - 1.5$.  \refone{The conformation is shown at five different times of the simulation, from green to dark blue (see colorbar), where only the final state is opaque.} Solid red line: trajectory of the center of mass. Red arrow: swimming direction. $\alpha$ is the angle between two fully extended configurations when $N_{\lambda} = 0$. See also Supplementary Movies.}
  \label{fig:traj_bacillaria}
\end{figure}

The wavelength of the deformation wave is varied in the range $N_\lambda = L_c/\lambda = 0 - 1.5$, where $N_\lambda$ is the number of wavelengths along the length colony, which amounts to setting $\Delta \phi = 0 - 2\pi/10$. Each simulation is run for  four sliding periods, $t_f = 4/f = 4T$, and is initialized from a rectangular stacked configuration along the $z-$axis where the rods are all parallel to the $x-$axis. 
We define $\tilde{V} = \Delta q_{COM}/L_c$ the number of chain lengths travelled per sliding period, where $\Delta q_{COM} =1/3\sum_{n=1}^3  \|\bq_{COM}((n+1)T) - \bq_{COM}(nT)\|$. 
Figure \ref{fig:traj_bacillaria} shows the center of mass  trajectories and \new{snapshots of the colony conformation}, for six typical values of $N_\lambda$ (see also Supplementary Movies \refone{generated with Matlab\textsuperscript{\textregistered}}).   When $N_\lambda  =   0$ the diatom chain expands and contracts symmetrically, leading to zero net velocity.  Interestingly,  the angle between two fully extended configurations is $\alpha \approx \pi/5$, which differs  dramatically from the value one would obtain without hydrodynamic interactions between the rods: $\alpha = \pi$.   When $N_\lambda>0$, the colony breaks Stokes's reversibility and exhibits a rich variety of trajectories and velocities that vary nonmonotonically. For $N_\lambda \leq 0.6$ the chain  travels in the $-x$ direction, which is the direction of the deformation wave, with some angle with respect to the $x-$axis set by hydrodynamics. A maximum in the swimming speed \new{is} obtained for $N_\lambda = 0.2$, for which the colony travels $64\%$ its lengths per sliding period. The net velocity goes back to zero for $N_\lambda \approx 0.6$, and switches to the positive $x-$direction, i.e.\ opposite to the wave, for $N_\lambda \approx 0.65$. Beyond that value, the colony conformation and the trajectories  are similar to the ones of a beating flagellum. The dimensionless speed reaches a secondary maximum  at $N_\lambda \approx 1$, for which the micro-swimmer travels $29\%$ of its length per sliding period. 
Altogether these preliminary results show that the swimming dynamics of \textit{Bacillaria Paxillifer} is rich, intricate and differs from flagellar dynamics  for which net motion is always opposite to the wave direction and optimal for $N_\lambda \approx 1$ \cite{Higdon1979b}. A more detailed study on the role of hydrodynamic interactions and the optimal swimming strategies in terms of efficiency \cite{Lighthill1975,Lauga2009} will be carried out in a forthcoming paper.

\subsection{Single bacterium}
\label{sec:bacterium}

Many unicellular organisms, like bacteria or protists, grow flagella that they use to self-propel \cite{Guasto2012, Nielsen2021}.
The flagellum is a flexible filament around $20\si{nm}$ in diameter and several micrometers in length \cite{Turner2000}.
When rotated by the molecular motors at their base, they take a helical form whose rotation allows the microorganims to swim \cite{Zhang2021}.
Under a constant angular velocity the flagellum reaches a steady stated determined by its flexibility.
When the flagellum is at the steady state it can be considered a rigid body as the bending forces are perfectly balanced by the fluid drag \cite{Berg1973,Block1989,Trachtenberg1992}.
For this reason many authors have modeled the flagella as rigid objects \cite{Shum2010,Shum2017,Higdon1979}, we follow this approach here
\refthree{and compare our results to those of Higdon, who employed a method of images and the slender body theory
  to calculate the swimming speed of bacteria  \cite{Higdon1979}.}

\begin{figure}
  \begin{center}
    \includegraphics[width=0.49 \columnwidth]{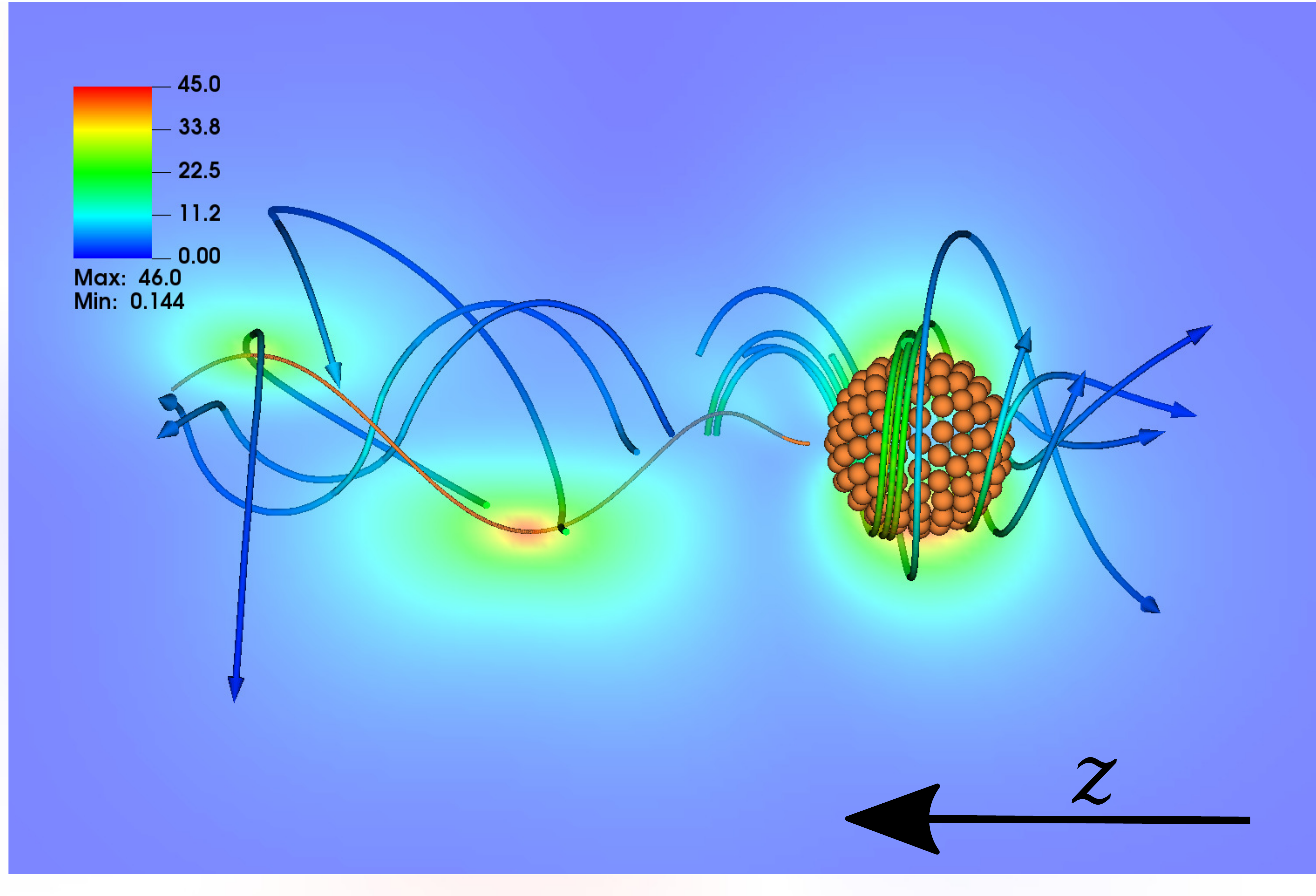}
    \includegraphics[width=0.49 \columnwidth]{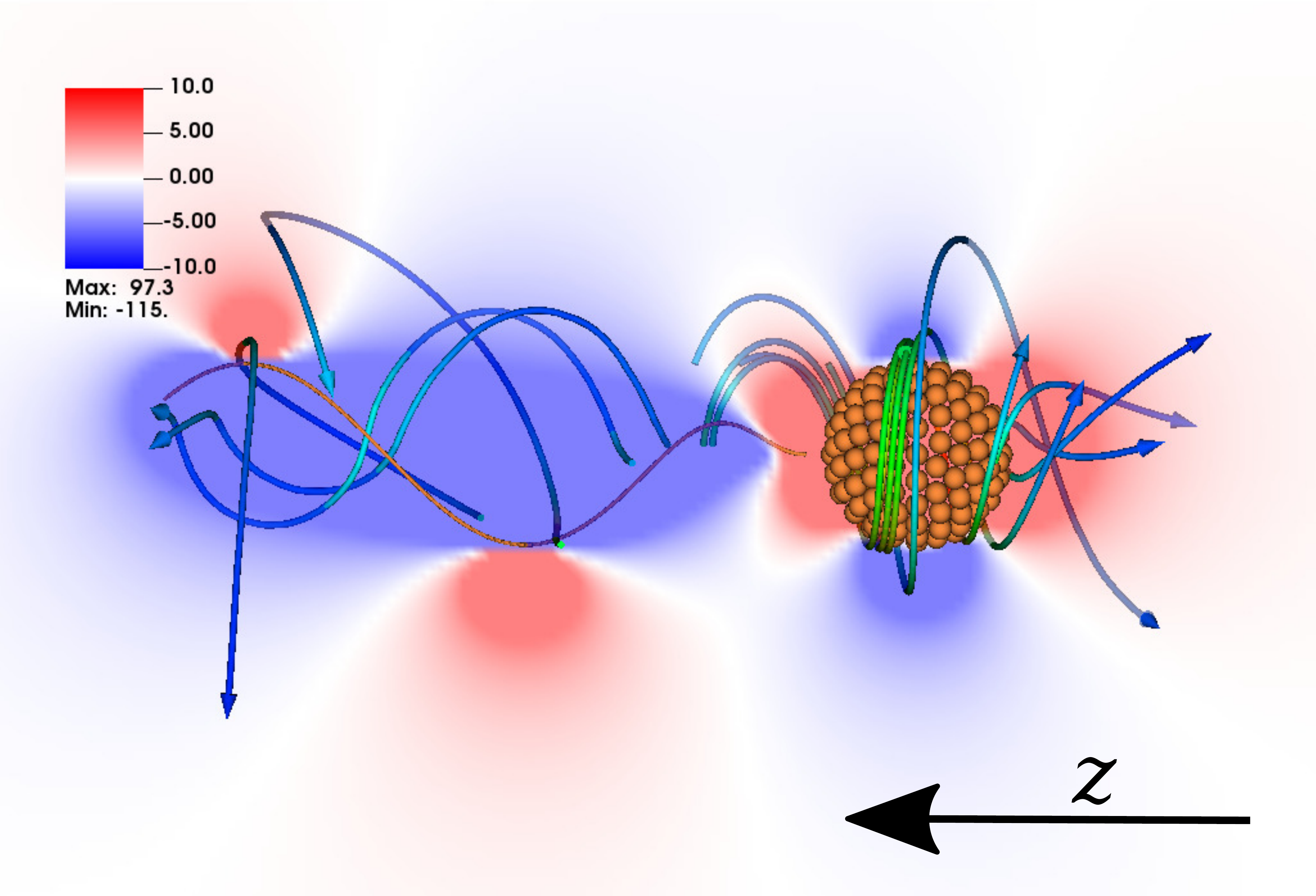}
    \caption{
      Streamlines around an active bacterium \refone{in free space} and slice of the velocity magnitude (left) and the vorticity \refthree{component along the $z$-axis} (right).
      \new{The bacterium has a body of radius $R=1\,\si{\mu m}$ and a slender flagellum with radius $h/R = 0.02$,
        length $L=10\,\si{\mu m}$ and $N_{\lambda}=1.14$ rotating at $\omega=100\,\si{Hz}$.}
    }
    \label{fig:bacteria_fields}
  \end{center}
\end{figure}

In this section we will use the model introduced in Section \ref{sec:time_convergence}  to study the dynamics of a single spherical bacterium with one  flagellum \refone{in free space}.
The flagellum assumes an helical shape where the equation of the centerline with respect to its attachment point ($z=0$) is given by \cite{Higdon1979,Smith2009,Ishimoto2019}
\eqn{
\bx(z) = \pare{\fr{1}{k} (1 - e^{-k^2z^2})\cos(kz),\,  \fr{1}{k} (1 - e^{-k^2z^2})\sin(kz), z}.
}
An example of a flagellum attached to a spherical bacterial body is shown in Fig. \ref{fig:bacteria_fields}.
In this section we study the speed of the bacterium for different flagellum lengths $L$ and  wavenumbers $k$.
We will see that the number of wavelengths in the flagellum, $N_{\lambda}=z_{\tex{max}}k/2\pi$, controls the speed \cite{Higdon1979}.

The flagellum can be rotated in two different ways: either by prescribing a constant angular velocity of the tail relative to the body, or by applying equal but opposite torques to the body and flagellum so that the whole bacterium is   force and torque-free. 
For a single bacterium swimming in unbounded space, both options provide equivalent results.
Here we fix the relative angular velocity between the spherical body and the flagellum.

\begin{figure}
  \begin{center}
    \includegraphics[width=0.8 \columnwidth]{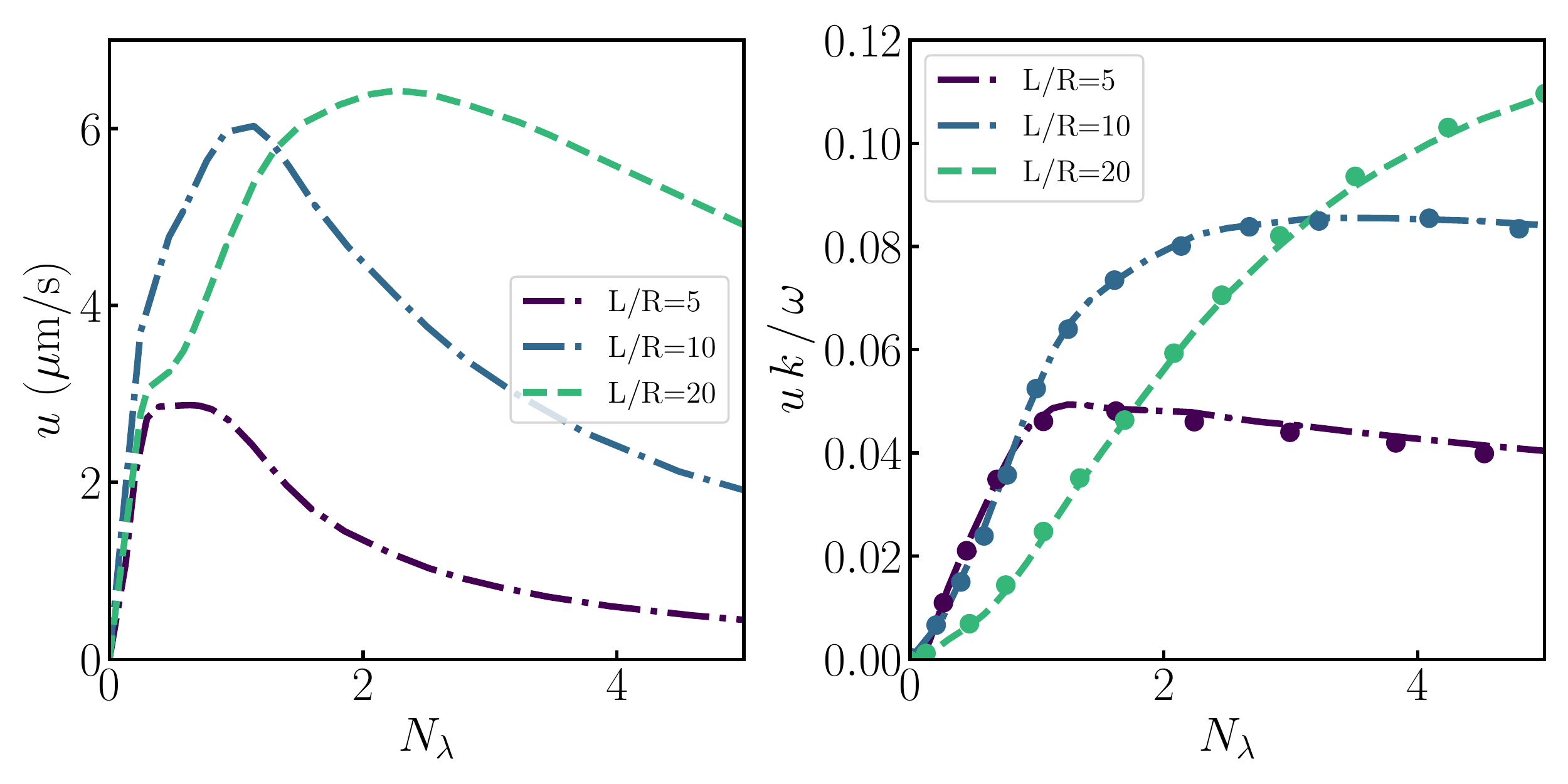}
  \caption{Swimming speed versus number of wavelengths along the flagellum $N_{\lambda}=z_{\tex{max}}/\lambda$
    for  three flagellum lengths $L$. 
    The radius (thickness) of the flagellum is \refthree{$h / R = 0.02$} and the angular velocity is \refthree{$\omega=100\, \si{s}^{-1}$}.
    On the right we normalize the swimming speed with the helical wave speed $\omega /k$, with $k$ the wavenumber of the helical wave,
    \refthree{and compare with the results of Higdon (symbols) \cite{Higdon1979}}.
  }
  \label{fig:bacteria_speed}
  \end{center}
\end{figure}

Figure \ref{fig:bacteria_fields} shows the streamlines around a bacterium together with the magnitude of the velocity (left) and vorticity fields (right).
It can be observed that the flagellum and the spherical body rotate in opposite directions, as required for a torque-free swimmer.
In the absence of walls or other swimmers, the bacterium swims with an average constant speed and direction.
In Figure \ref{fig:bacteria_speed} we plot the swimming speed  for different
flagellum lengths, $L$, and different number of wavelengths $N_{\lambda}=z_{\tex{max}}/\lambda$.
As in the seminal work of Higdon \cite{Higdon1979} the absolute speed is maximized for $N_{\lambda} \approx 1$, i.e.\ when the helical wave does a full turn along the flagellum.
\refthree{When normalized by the wave speed, $\omega/k$, the motion of the flagellum is transmitted better for larger values of $N_{\lambda}$.
Our results agree well with those of Higdon \cite{Higdon1979}.
}

\subsection{Bacterial suspension}
\label{sec:many_bacteria}

To demonstrate the capabilities of our method we simulate a suspension with $N=100$ bacteria swimming above a no-slip wall.
\refthree{Most numerical studies on bacterial suspensions use either continuum formulations \cite{Saintillan2008a, Lushi2016},
  minimal models based on regularized Stokes dipoles \cite{Lushi2014} or methods such as Lattice Boltzmann or Multiparticle Collision Dynamics
  that work at finite Schmidt and Reynolds numbers \cite{Scagliarini2022, Qi2022}.
  Instead, we solve the Stokes flow around the whole bacteria, each discretized  with $200$-blobs (body + flagellum)}.
  This is a large system with $N\times 200 = 20, 000$ blobs and $61, 800$ unknowns.
  We consider two types of boundary conditions: a semi-infinite system only bounded by the wall and a system with periodic boundary conditions (PBC)
  in the directions parallel to the wall.
  We include a short range steric repulsion between the bacteria and between the bacteria with the wall to prevent overlaps.
  To compute the hydrodynamic interactions between the blobs above a no-slip wall we use a Fast Multipole  implementation of the wall-corrected RPY tensor, available in the library STKFMM \cite{Yan2018a, Yan2020}, that enables PBC along the lateral directions.
  In both cases the computational cost scales quasilinearly with the total number of blobs which allows to run long simulations.
  The tolerance of the linear and nonlinear solvers are set to $\epsilon = 10^{-6}$ and $\delta = 10^{-10}$ respectively.
  The angular velocity of the flagellum is set to $\omega=100\, \si{Hz}$ and the time step to $\dt=0.005\, \si{s}$, which corresponds to $12$ steps per flagellum rotation.
We simulate the system for  $2500$ time steps, which takes about $12$h on a $28$-core computer.

\begin{figure}
  \begin{center}
    \includegraphics[width=1.0 \textwidth]{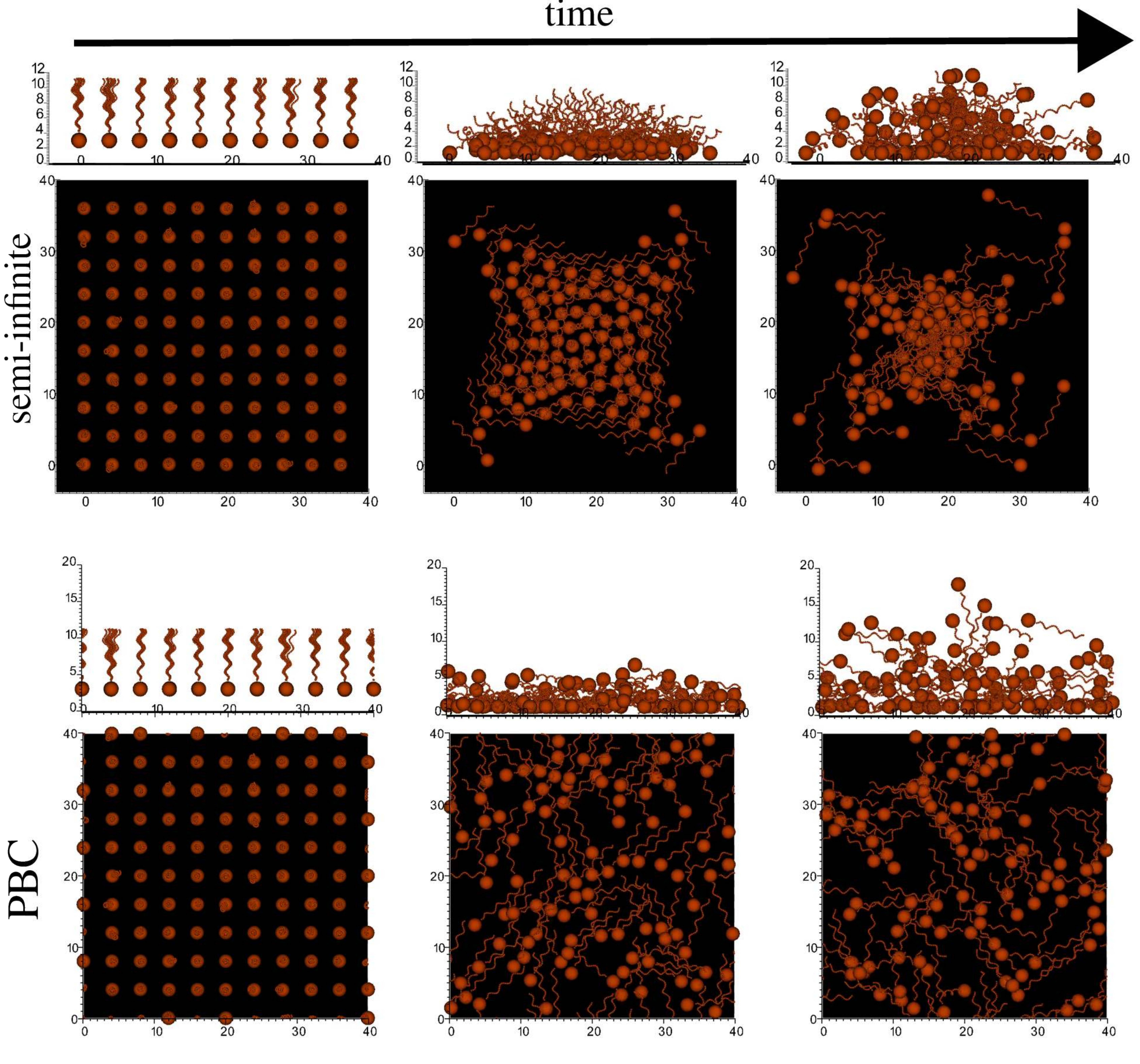}
    \caption{Top and lateral views of a suspension of $N=100$ bacteria above a no-slip wall at times $0$, $5$ and $12\,\si{s}$.
      Top row:  semi-infinite  domain only bounded by the wall.
      Bottom row: periodic boundary conditions (PBC) in the directions parallel to the wall.
      See also movies in the supplemental material. 
    }
    \label{fig:bacteria_suspension}
  \end{center}
\end{figure}

Figure \ref{fig:bacteria_suspension} shows a few snapshots of the bacterial suspension, see also Supplementary Movies \refone{generated with VisIt \cite{Childs2012}}.
Initially, the bacteria form a square lattice and are oriented towards the wall with small perturbations in their orientations. 
They start swimming towards the wall but the monolayer soon becomes unstable and forms dense clusters.
In the unbounded domain a single cluster is formed, while with PBC the suspension exhibits more complex patterns.
\refone{Additionally, in the unbounded domain the cluster preserves the tendency of bacterium to swim in circles near walls \cite{Maeda1976} due to the torque dipole
created by the opposite rotation of body and flagellum and the hydrodynamic interactions with the wall \cite{Lauga2006}.
With PBC the strong interactions between neighbor bacteria suppress this effect.}
At longer times the flow created by the bacteria destroys the monolayer: some bacteria reorient away from the wall, while others stay tilted towards it.

The collective motion that destabilizes the initial lattice is driven by the flow generated by the microswimmers  \cite{Saintillan2008a}.
The flow induced by a bacterium can be modeled, to a low order approximation, as an extensile force dipole as the flagellum pushes the fluid backwards while the
bacterium body pushes it forward.
\refone{At the same time the fluid is pulled towards the bacterium from its sides to maintain the incompressibility condition.
  This flow causes a lateral attraction between swimmers, creating clusters and eventually destabilizing the monolayer.}
Suspensions of swimmers with this characteristic flow signature, known as \emph{pushers}, exhibit a complex dynamics with instabilities and large scale density fluctuations
\cite{Dombrowski2004, Saintillan2007}.
Interestingly, it is well known that many microswimmers, such as bacteria, are hydrodynamically attracted to obstacles and surfaces
\cite{Berke2008,Spagnolie2012,Mousavi2020}.
Such interactions, if strong enough, could stabilize a monolayer of microswimmers near a wall.
In this particular experiment the destabilizing effect of the pushers' flow dominates over the hydrodynamic attraction to the wall.
Our model, which includes both effects as well as the steric interactions and collisions between bacteria, opens a window to explore the stability
of microswimmer suspensions beyond continuum model approximations.


\section{Conclusions}
We have presented a new framework to simulate large suspensions of articulated bodies.
Our velocity formulation of the constraints between bodies enables to write the mixed resistance-mobility problem  as a single linear system.
We solved this linear system with a preconditioned iterative solver that couples effectively with any numerical method to compute hydrodynamic interactions between rigid bodies.
Interestingly, the solver convergence is independent of the system size and constraint type, therefore  allowing to simulate large suspensions in a scalable fashion.
Combining our new reconstruction method, which generalizes the robot-arm model, with a simple and costless correction procedure, one can track articulated bodies with a reduced number of degrees of freedom while avoiding constraint violations. 
Our method is robust and flexible, it applies to a wide variety of physical and biological systems. We have illustrated some of them involving deformable filaments, shells, and various types of micro-swimmers. Our implementations are freely and publicly available   so that others can use its capacity to address new problems.
Some applications of interest happen at very small scales, such as the motion of bacteria, membranes, lipid bilayers,  actin filaments or molecular motors, where thermal fluctuations play an important role. 
One future direction is, therefore, to include Brownian motion in our framework \cite{Morse2004}.

\section*{Acknowledgments}
F.B.U.\ acknowledges support from ``la Caixa'' Foundation (ID 100010434), fellowship LCF/BQ/PI20/11760014, and from the European Union's Horizon 2020 research and innovation programme under the Marie Skłodowska-Curie grant agreement No 847648
and also by the Basque Government through the BERC 2022-2025 program and by the Ministry of Science, Innovation and Universities:
BCAM Severo Ochoa accreditation SEV-2017-0718.
B.D.\ acknowledges support from the French National Research Agency (ANR), under award ANR-20-CE30-0006.  B.D.\ also thanks the NVIDIA Academic Partnership program
for providing GPU hardware for performing some of the simulations reported here.

\appendix
\section{No net constraint force and torque on articulated bodies}
\label{sec:convervation_momentum}
We prove here that the constraint forces do not exert any total force or torque on the articulated  body. We consider again the linear system \eqref{eq:linear_system_general},
\eqn{
\left[\begin{array}{cc} 
\new{-}\bN \bC^T& \bI  \\
 \bzero & \bC 
\end{array} \right]
\left[\begin{array}{c}
\bphi \\
\bU 
\end{array} \right] =
\left[\begin{array}{c}
\bN \bF \\
\bB
\end{array} \right],
}
The block of the matrix $\bC$ corresponding to the link $n$ connecting bodies $p$ and $q$ is given by
\eqn{\bC_n = \left[\bzero \cdots\;\; \bI\;\; -\Delta\bl_{np}^\times\;\;\bzero \cdots\;\;-\bI\;\; \Delta \bl_{nq}^\times\;\;\bzero \cdots\right],}
and the constraint forces and torques exerted by link $n$ on those two bodies are
\eqn{\bF^C = \bC_n^T \bphi=\pare{\bzero \cdots\;\;\bphi_n\;\; \Delta \bl_{np}\times \bphi_n\;\;\bzero \cdots\;\; -\bphi_n\;\; -\Delta \bl_{nq}\times \bphi_n\;\;\bzero \cdots\;\;}.}
The total force on the articulated body exerted by the constraint is therefore $\bF^C_T=\bF^C_p+\bF^C_q=\bphi_n - \bphi_n=\bzero$. The total torque, here defined around the body $p$, is 
\eqn{\bT^C_T &= \bT^C_p + \bT^C_q + (\bq_q - \bq_p)\times \bF^C_q \nonumber \\
&= \left(\bq_p + \Delta \bl_{np} - \bq_q - \Delta \bl_{nq}\right) \times \bphi_n \nonumber \\ & = \bg_n\times \bphi_n= \bzero.}
\new{These} results apply to constant and time dependent constraints.

\section{Jacobian of the objective function}
\label{sec:Jacobian}
In this appendix we write the expression for the Jacobian of the residual \eqref{eq:min_probl_a}.
To derive the Jacobian it is enough to focus on  Eq. \eqref{eq:residual} since all the constraints have the same form.
To further simplify the equations we drop subscripts and derive the Jacobian for the shorter equation
\eqn{
  \label{eq:residual_simple}
  \bg(\bdq,\bdth) = \bq + \bdq  + \bR(\bdth \bullet \btheta) \bDl.
}
Note that the variables of the two rigid bodies, $p$ and $q$, enter in \eqref{eq:residual} in the same way with a plus or minus sign.
Therefore the full Jacobian of \eqref{eq:residual} will have just twice as many nonzero terms, each half multiplied by $+1$ or $-1$.
Using the property of the rotation matrices $\bR(\btheta_2 \bullet \btheta_1)=\bR(\btheta_2)\bR(\btheta_1)$, we can write the residual as
\eqn{
  \bg(\bdq,\bdth) = \bq + \bdq  + \bR(\bdth) \wtil{\bDl}.
}
where $\wtil{\bDl} = \bR(\btheta) \bDl$.
Finally, the rotation matrix associated to the unit quaternion $\btheta=(s, \bp)$ is \cite{Delong2015b}
\eqn{
  \bR = 2 \corchete{\bp \bp^T + s \bp^{\times} + \pare{s^2 - \fr{1}{2}} \bI}, 
  \label{eq:rot_mat}
}
where $\bp^{\times} \bx = \bp \times \bx$ for any vector $\bx$.
Then, the constraint in indicial notation is
\eqn{
  \label{eq:residual_inidicial}
  g_i = q_i + \delta q_i + 2 p_i (p_j \wtil{\Delta l}_j) + s \epsilon_{ijk} p_k \wtil{\Delta l}_j + \pare{s^2 -\fr{1}{2}} \wtil{\Delta l}_i.
}
where $\epsilon_{ijk}$ is the Levi-Civita symbol.
From this expression it is easy to compute the elements of the Jacobian
\eqn{
  J_{i q_j} &= \fr{\partial g_i}{\partial q_j} =   \delta_{ij}, \\
  J_{i s} &=  \fr{\partial g_i}{\partial s} = \epsilon_{ijk} p_k \wtil{\Delta l}_j + 2s \wtil{\Delta l}_i \\
  J_{i p_l} &=  \fr{\partial g_i}{\partial p_l} = 2 \delta_{il} (p_j \wtil{\Delta l}_j) + 2 p_i \wtil{\Delta l}_l + s \epsilon_{ijl} \wtil{\Delta l}_j.
}
This Jacobian has dimensions $3 \times 7$, the Jacobian of the full nonlinear equation \eqref{eq:min_probl_a} will have
dimensions $3P \times 7M$ for $M$ rigid bodies linked by $P$ constraints, however, only $30 P$ elements will be nonzero and thus the Jacobian will be sparse.






\section{Midpoint scheme}
\label{app:midpoint}
Below we outline the correction algorithm embedded in an Explicit Midpoint scheme (or RK2).\\

\noindent Time loop: for $k = 0,..,N_{it}-1$
\begin{enumerate}
\item Solve the linear system \eqref{eq:linear_system} at $t_k = k\Delta t$ to obtain the body velocities $\bU^{(k)}$,
  
\item Compute the translational velocity of the COM's  $\bu_{COM}^{(k)}$ using \eqref{eq:ucom},
  
\item Update the COM  and the orientations of each assembly to the midpoint $t_{k+1/2} = (k+1/2)\Delta t$
  \begin{eqnarray}
    \bq_{COM}^{\star} &=& \bq_{COM}^{(k)} + \frac{1}{2}\Delta t \bu^{(k)}_{COM}\\
    \btheta^{\star} &=& \btheta_{\bomega^{(k)}\frac{1}{2}\Delta t}\bullet\btheta^{(k)},
  \end{eqnarray}
  
\item Reconstruct the articulated bodies using the pseudo-inverse \eqref{eq:pseudo-inv-sol}, where  $\bz^a = \bz^a(\btheta^\star,t_{k+1/2})$   and then apply Eq.\  \eqref{eq:reconst-up-COM} to obtain the body positions around the updated centers of mass $\bq^{\star}$.
  
\item Evaluate the updated constraint vector $\bg^\star = \bg(\bq^\star,\btheta^\star,t_{k+1/2})$ and  check the constraint violation:\\ if $\|\bg^\star\|^2_2 < \delta$ then 
  \eqn{
    \bq^{n+1/2} \leftarrow \bq^{\star}\\
    \btheta^{n+1/2} \leftarrow \btheta^{\star}
  }
  else
  \begin{itemize}
  \item Solve the minimization problem \eqref{eq:min_probl_a}-\eqref{eq:min_probl_b}  with the nonlinear solver to find the increments $(\breve{\bdq}, \breve{\bdth})$ 
  \item Correct the body positions and orientations
    \begin{eqnarray}
      \bq^{n+1/2} &=& \bq^{\star} + \breve{\bdq},\\
      \btheta^{n+1/2} &=& \breve{\bdth}\bullet\btheta^{\star}.
    \end{eqnarray}
  \end{itemize}
  
\item Solve the linear system \eqref{eq:linear_system} at the midpoint  $t_{k+1/2} = (k+1/2)\Delta t$ to obtain the body velocities $\bU^{(k+1/2)}$,
  
\item Compute the translational velocity of the COM's  $\bu_{COM}^{(k+1/2)}$ using \eqref{eq:ucom},
  
\item Update the COM   and the orientations of each assembly to the next time step
  \begin{eqnarray}
    \bq_{COM}^{\star} &=& \bq_{COM}^{(k)} + \Delta t \bu^{(k+1/2)}_{COM}\\
    \btheta^{\star} &=& \btheta_{\bomega^{(k+1/2)}\Delta t}\bullet\btheta^{(k)},
  \end{eqnarray}
  
\item Reconstruct the articulated bodies using the pseudo-inverse \eqref{eq:pseudo-inv-sol}, where  $\bz^a = \bz^a(\btheta^\star,t_{k+1})$    and then apply Eq.\  \eqref{eq:reconst-up-COM} to obtain the body positions around the updated centers of mass $\bq^{\star}$.
  
\item Evaluate the updated constraint vector $\bg^\star = \bg(\bq^\star,\btheta^\star,t_{k+1})$ and  check constraint violation:\\ if $\|\bg^\star\|^2_2 < \delta$ then 
  \eqn{
    \bq^{n+1} \leftarrow \bq^{\star}\\
    \btheta^{n+1} \leftarrow \btheta^{\star}
  }
  else
  \begin{itemize}
  \item Solve the minimization problem \eqref{eq:min_probl_a}-\eqref{eq:min_probl_b} with the nonlinear solver to find the increments $(\breve{\bdq}, \breve{\bdth})$ 
  \item Correct the body positions and orientations
    \begin{eqnarray}
      \bq^{n+1} &=& \bq^{\star} + \breve{\bdq},\\
      \btheta^{n+1} &=& \breve{\bdth}\bullet\btheta^{\star}.
    \end{eqnarray}
  \end{itemize}
  
\end{enumerate}

\bibliographystyle{elsarticle-num}
\bibliography{Biblio.bib}

\end{document}